\documentclass[preprint,eqsecnum,aps]{revtex4-1}
\usepackage{graphicx}
\newcommand{\B}{{\bf B}}
\newcommand{\E}{{\bf E}}
\newcommand{\F}{{\bf F}}
\newcommand{\ct}{{\cal T}}
\newcommand{\bh}{\hat{\bf B}}
\newcommand{\bta}{{\mbox{\boldmath{$\theta$}}}}

\newcommand{\br}{{\bf r}}

\newcommand{\bac}{{\bf a}}
\newcommand{\C}{{\cal C}}

\newcommand{\p}{{\bf p}}
\newcommand{\dt}{{\Delta t}}

\newcommand{\ta}{{\theta}}

\newcommand{\bi}{\hat{\bf x}}
\newcommand{\bj}{\hat{\bf y}}
\newcommand{\bv}{{\bf v}}
\newcommand{\bu}{{\bf u}}
\newcommand{\bk}{\hat{\bf z}}

\newcommand{\w}{\omega}
\newcommand{\e}{{\rm e}}

\newcommand{\nn}{\nonumber}

\newcommand{\la}{\label} 
\newcommand{\be}{\begin{equation}}
\newcommand{\ee}{\end{equation}}
\newcommand{\ba}{\begin{eqnarray}}
\newcommand{\ea}{\end{eqnarray}}
\begin{document}
\title{The anatomy of Boris type solvers and the Lie operator formalism for
deriving large time-step magnetic field integrators}
\author{Siu A. Chin$^*$ and Durward Cator}
\affiliation{Department of Physics and Astronomy, Texas A\&M University,
	College Station, TX 77843, USA}

\begin{abstract}	
	
This work gives a Lie
operator derivation of various Boris solvers via 
a detailed study of trajectory errors in a constant magnetic field. 
These errors in the gyrocenter location and the gyroradius are the foundational basis
for why Boris solvers existed, independent of any finite-difference schemes.
This work shows that there are two distinct ways of eliminating these errors so that
the trajectory of a charged particle in a constant magnetic field is exactly on
the cyclotron orbit. One way reproduces the known second-order symmetric Boris solver.
The other yields a previously unknown, but also on-orbit solver, not derivable from finite-difference schemes. 
By revisiting some historical calculations, it is found that many publications do not 
distinguish the poorly behaved leap-frog Boris solver from the symmetric second-order 
Boris algorithm. This symmetric second-order Boris solver's trajectory is much more accurate and 
remains close to the exact orbit in a combined {\it nonuniform} electric and magnetic field
at time-steps greater than the cyclotron period. Finally, this operator formalism
showed that Buneman's cycloid fitting scheme is mathematically identical to Boris' on-orbit solver 
and that Boris' E-B splitting is unnecessary. 

\bigskip
\bigskip
\bigskip
\bigskip
\bigskip
\bigskip


\bigskip
\bigskip
\bigskip
\bigskip
\bigskip
\bigskip
\bigskip
\bigskip
\bigskip
\bigskip

\noindent
$^*$ Corresponding author, chin@physics.tamu.edu

\end{abstract}
\maketitle

\section {Introduction}
\la{in}

The leap-frog (LF) Boris (or Buneman-Boris) solver\cite{bun67,bor70} has been widely used
in plasma physics simulations\cite{bir85,par91,vu95,sto02,vay08,qin13,he15,zen18,rip18,ric20}
for decades. Yet, from the very beginning, there are continued disagreements as to the nature
of Boris' E-B splitting solver and Buneman's drift-subtracting scheme. Buneman has claimed\cite{bun67} that
by ``cycloid-fitting'', his scheme is exactly on-orbit in a constant electric and magnetic field,
while Boris\cite{bor70} (and Ref.\onlinecite{bir85}) has explicitly stated that his solver is not. 
However, in the literature, there remain intermittent claims\cite{sto02,ric20} that the ``Boris'' 
solver is also exactly on-orbit.

In this work, we use the Lie operator formalism\cite{chin08} to derive Boris-type algorithms 
(to be precisely defined in Sect.\ref{diff})
from first principle, independent of any finite-difference schemes. In a constant
magnetic field, this formalism's two first-order algorithms have three basic errors,
the off-center coordinates of the gyro-circle and its radius. Correcting these three errors
then defines various Boris solvers. This is the most fundamental characterization of a Boris solver, 
independent of its historical tie to the implicit midpoint method\cite{bun67,bor70} and
its distinctive Cayley\cite{kna15,he15} (or Crank-Nicolson\cite{ric20}) form of rotation.
   
Leap-frog type algorithms, which update the position and momentum
variables sequentially, were historically novel as compared to Runge-Kutta
type algorithms, which update variables synchronously. However, the rise of modern 
symplectic integrators\cite{dep69,dra76,yos93,ser16} (SI), which identified\cite{chin20} sequential 
updating as the distinguishing hallmark of canonical transformations, has made sequential 
updating the new norm for classical dynamics algorithms. 
While Runge-Kutta schemes are generally not phase-volume preserving, 
as will be shown in the next Section, {\it any} sequential updating of 
the position and momentum variables is automatically volume preserving.
This work shows that the original LF Boris solver, despite being second-order 
by having {\it symmetric initial positions}, shares the same large error gyroradius as 
first-order sequential algorithms and is not on-orbit. 
The on-orbit solver is the {\it intrinsically  symmetric} second-order Boris solver
having the correct gyroradius. The subtle differences between these
two solvers are carefully explained in Sect.\ref{diff}.
By revisiting some historical calculations\cite{par91}, 
it is found that the much more accurate symmetric Boris solver has not been used for 
large time step simulations. More recent publications\cite{he15,zen18,rip18,ric20}
also do not distinguish these as two different algorithms.

We begin by reviewing the Lie operator (or series) method\cite{dep69,dra76,yos93,ser16} 
of deriving magnetic integrators\cite{chin08} in Sect.\ref{om}. 
This same formalism is used to derive symplectic integrators, except 
that by using the mechanical momentum in the Lorentz force law rather than the canonical momentum
in the Hamiltonian, the resulting algorithms are Poisson\cite{kna15} integrators, rather than symplectic. 
To make the derivative Lie operators more comprehensible, we introduce the
cross-product operator $\C$ so that the velocity update in a magnetic field can be 
immediately recognized as a rotation. Later in Sect.\ref{seceb}, the operator $\C$ will simplify 
discussions on Boris' original inversion algorithm and the norm preserving 
Cayley\cite{kna15,he15} (or Crank-Nicolson\cite{ric20}) approximation of the exponential.

In Sect.\ref{first}, we derive two basic first-order magnetic field algorithms
and scrutinize their trajectory errors for a constant magnetic field. 
Two distinct choices of eliminating these errors result in two different types of Boris solvers.
The conventional LF Boris solver corresponds to an intermediate form 
of the two first-order algorithms. In revisiting some historical large time step calculations\cite{par91},
it is found that only one of the first order solvers is used, which is not even the Boris LF solver.

In Sect.\ref{sec}, intrinsically symmetric, on-orbit, second-order Boris solvers are derived, 
including one that is not derivable from finite-difference schemes. The difference
between the symmetric and the LF Boris solver, together with a simple explanation of 
their distinct gyro-radii, are given in Sect.\ref{diff}. 

In Sect.\ref{seceb}, with the inclusion of
the electric field, we show that the velocity update
of Boris' E-B splitting method is
mathematically identical to Buneman's drift-subtracting scheme,
which in term, is exactly the same as that given by  
the Lie operator formalism. The Boris E-B splitting is of practical convenience, 
but unnecessary in that there is no defect in 
Buneman's scheme that is solved by the splitting.
By repeating some historical\cite{par91} and more recent\cite{he15} calculations, 
it is again found that the intrinsic symmetric second-order Boris solver is the best 
trajectory tracker at large time steps. Conclusions are drawn in Sect.\ref{con}.

\section {The Lie operator method}
\la{om}

The equations of motion for a charged particle in a static electric $\E(\br)$ and 
magnetic field $\B(\br)=B(\br)\bh(\br)$ can be written as
\be
\frac{d\br}{dt}=\bv\quad{\rm and}\quad \frac{d\bv}{dt}=\w(\br)\bh(\br)\times\bv+\bac(\br)
\la{em}
\ee
where $\bv\equiv\p/m$, $\w(\br)=(-q)B(\br)/m$ and $\bac(\br)=\F(\br)/m=q\E(\br)/m$.
(Note that $\w>0$ when $q<0$, since the cyclotron motion of a negatively charge particle is
counter-clockwise when the magnetic field is out of the page.)
The vectors $\br$ and $\bv$ are fundamental and independent dynamical variables. 

For any other dynamical variable $W(\br,\bv)$, its evolution through (\ref{em}), is given
by
\ba
\frac{dW}{dt}&=&\frac{\partial W}{\partial \br}\cdot\frac{d\br}{dt}
+\frac{\partial W}{\partial \bv}\cdot\frac{d\bv}{dt}\nn\\
&=&\Bigl(\bv\cdot\frac{\partial}{\partial \br}
+(\w\bh\times\bv+\bac)\cdot\frac{\partial}{\partial \bv}\Bigr)W,
\la{evolm}
\ea
which can be directly integrated to yield the operator solution
\be
W(t)=\e^{t(T+V_{BF})}W(0)
\la{gen}
\ee
where one has defined Lie operators\cite{dep69,dra76,yos93}
\be
T=\bv\cdot \frac{\partial}{\partial \br}
\la{magt}
\ee
and
\ba
V_{BF}
&=&\w(\bh\times\bv)\cdot\frac{\partial}{\partial \bv}+\bac\cdot\frac{\partial}{\partial \bv}
\equiv V_B+V_F.
\la{magv}
\ea
To solve (\ref{gen}), one takes $t=n\dt$ so that (\ref{gen}) can be reduce to 
$n$ iterations of the short-time operator $\exp[\dt(T+V_{BF})]$ via
Baker-Campbell-Hausdorff type approximations, of which the simplest examples are
\ba
\e^{\dt(T+V_{BF})}&\approx&\e^{\dt T}\e^{\dt V_{BF}}\nn\\
&\approx& \e^{\dt T}\e^{\dt V_B}\e^{\dt V_F}. 
\ea
The action of each individual operator can easily be computed via series expansion:
\ba
\e^{\dt T}
\left(
\begin{array}{c}
	\br\\
	\bv
\end{array}
\right)
&=&(1+\dt\bv\cdot\frac{\partial}{\partial \br}
+\frac{\dt^2}{2}(\bv\cdot\frac{\partial}{\partial \br})^2+\cdots)
\left(
\begin{array}{c}
	\br\\
	\bv
\end{array}
\right)\nn\\
&=&
\left(
\begin{array}{c}
	\br+\dt\bv\\
	\bv
\end{array}
\right)
\la{expt}
\ea
\ba
\e^{\dt V_F}
\left(
\begin{array}{c}
	\br\\
	\bv
\end{array}
\right)
&=&(1+\dt\bac\cdot\frac{\partial}{\partial \bv}
+\frac{\dt^2}{2}(\bac\cdot\frac{\partial}{\partial \bv})^2+\cdots)
\left(
\begin{array}{c}
	\br\\
	\bv
\end{array}
\right)\nn\\
&=&
\left(
\begin{array}{c}
	\br\\
	\bv+\dt\bac
\end{array}
\right).
\la{expvf}
\ea
More generally, the product approximation
\be
\e^{\dt(T+V_F)}=\prod_{i=1}^{N}\e^{a_i\dt T}\e^{b_i\dt V_F}
\la{sym}
\ee 
with suitable coefficients $a_i$ and $b_i$, then generates sequential updates
(\ref{expt}) and (\ref{expvf}), which is a {\it symplectic integrator}\cite{yos93,ser16,chin20} 
of arbitrary order for
solving (\ref{em}) without a magnetic field. Since (\ref{expt}) and (\ref{expvf}) are
sequential translations, it is obvious that any algorithm of the form (\ref{sym}) 
is phase-volume preserving.
Likewise, the approximation
\be
\e^{\dt(T+V_B)}=\prod_{i=1}^{N}\e^{a_i\dt T}\e^{b_i\dt V_B}
\la{mag}
\ee
will generate sequential updates which are {\it exact energy conserving}\cite{chin08} for solving
(\ref{em}) with only a magnetic field. This Lie operator method is powerful in that even
without knowing the explicit form of $\exp(b_i\dt V_B)$, one can prove that (\ref{mag}) is
exact energy conserving because by (\ref{expt}),
\be
\e^{a_i\dt T}\bv^2=\bv^2
\ee 
and 
\be
\e^{b_i\dt V_B}\bv^2=(1+b_i\dt\w(\bh\times\bv)\cdot\frac{\partial}{\partial \bv}
+\cdots)\bv^2=\bv^2,
\la{vpre}
\ee   
since after differentiating $\bv^2$, the resulting triple product vanishes.
Moreover (\ref{vpre}) implies that the effect of $\e^{b_i\dt V_B}$ on $\bv$ 
must only be a rotation. Consequently
the algorithm (\ref{mag}) is a sequence of translations and rotations and therefore
again phase-volume preserving. 
Finally, the approximation
\be
\e^{\dt(T+V_{BF})}=\prod_{i=1}^{N}\e^{a_i\dt T}\e^{b_i\dt V_{BF}}
\la{emag}
\ee
will generate sequential updates of a {\it Poisson integrator}\cite{kna15,chin08} for solving 
charged particle trajectories in a combined electric and magnetic field
to arbitrary precision. Again, even without knowing the explicit form of
$\exp(\dt V_{BF})\bv$ it is easy to prove that (\ref{emag}) must also be
volume preserving. This is because $\exp(\dt V_{BF})$ itself can be approximated
to any order of accuracy as 
\be
\e^{\dt V_{BF}}=\prod_{i=1}^{N}\e^{a_i\dt V_B}\e^{b_i\dt V_F}
\la{embf}
\ee
which is only a sequence of translations and rotations and therefore must be 
phase-volume preserving.

However, the explicit form of $\exp(\dt V_B)\bv$ and $\exp(\dt V_{BF})\bv$
are known from Ref.\onlinecite{chin08},
\begin{equation}
	\e^{\dt V_B}
	\left(\begin{array}{c}
		\br \\
		\bv
	\end{array}\right)
	=
	\left(\begin{array}{c}
		\br\\
		\bv_B(\br,\bv,\dt)
	\end{array}\right)
	\label{vbeq}
\end{equation}
\begin{equation}
	\e^{\dt V_{BF}}
	\left(\begin{array}{c}
		\br \\
		\bv
	\end{array}\right)
	=
	\left(\begin{array}{c}
		\br\\
		\bv_B(\br,\bv,\dt)+\bv_F(\br,\bv,\dt)
	\end{array}\right)
	\label{vbfeq}
\end{equation}
where 
\ba
\bv_B(\br,\bv,\dt)&=&\bv+\sin\theta(\bh\times\bv) +
(1-\cos\theta)\bh\times(\bh\times\bv)\nn\\
&=&\bv_\Vert+\cos\theta\bv_\perp+\sin\theta(\bh\times\bv_\perp)\la{vb}\\
\bv_F(\br,\bv,\dt)&=&\dt\bac+\frac{1}{\w}(1-\cos\theta)\bh\times\bac+
\dt(1-\frac{\sin\theta}\theta )\bh\times(\bh\times\bac)\la{va}\\
&=& \dt\bac_\Vert 
+\frac{1}\w \Bigl[ (1-\cos\theta)(\bh\times\bac_\perp)
+ \sin\theta\bac_\perp\Bigr],
\la{vap}
\ea
with $\ta=\w(\br)\dt$ and where $\bv_{||}+\bv_{\perp}=\bv$, $\bac_{||}+\bac_{\perp}=\bac$
are components parallel and perpendicular to the local magnetic field direction $\bh(\br)$.

Eq.(\ref{vb}) is the effect of $\e^{\dt V_B}$ acting on $\bv$, which is to rotate only $\bv_{\perp}$ 
thereby preserving $|\bv|$ and the kinetic energy. This is the same as exactly solving
\be
\frac{d\bv}{dt}=\w(\br)\bh(\br)\times\bv
\la{beq}
\ee
holding $\br$ fixed. Define the cross-product operator $\C=(\bh(\br)\times\ )$.
Since 
\be
\C^2\bv_\perp=-\bv_\perp,
\ee
$\C$ behaves as $\sqrt{-1}=i$ when acting on any vector $\bv_\perp$
perpendicular $\bh(\br)$. The solution to (\ref{beq}) for time $\dt$ is therefore
\ba
\bv(\dt)&=&\e^{\dt\w\C}\bv
=\e^{\ta\C}(\bv_{||}+\bv_{\perp}),\nn\\
&=&\bv_{||}+\cos\ta\bv_\perp+\sin\ta\C\bv_\perp,
\la{ceul}
\ea
which is the same as (\ref{vb}) because $V_B$ is equivalent to $\C$
when acting on $\bv$:
\be
V_{B}\bv=\C\bv\quad{\rm and}\quad
(V_{B})^n\bv=\C^n\bv.
\ee
(Note that $\e^{\ta\C}=\cos\ta+\sin\ta\C$ when acting on $\bv_\perp$ in (\ref{ceul}) is just
Euler's formula $\e^{i\ta}=\cos\ta+i\sin\ta$ in disguise.)
This Lie operator method of exactly solving (\ref{beq}) is to be contrasted with finite
difference schemes, which have no means of solving (\ref{beq}) exactly without {\it ad hoc} adjustments.

Similar operator exponentiation\cite{chin08} results
(\ref{vap}), accounting for the $\E\times\B$ drift, will be discussed in Sect.\ref{seceb}.
The program is therefore complete for the generation of arbitrarily accurate magnetic field
integrators (\ref{mag}) or (\ref{emag}) in the limit of small $\dt$. 
The goal of this work, however is to show how Boris solvers can also be derived from 
this powerful machinery and to seek accurate integrators for solving magnetic field trajectories
at {\it large} $\dt$.

\section {First-order magnetic field integrators}
\la{first}

Boris\cite{bor70} originally derived his solver by modifying the implicit midpoint method.
Here, we will show how Boris type solvers can be derived systematically from the
Lie operator method without referencing any finite-difference schemes.
By Boris type solver, we shall mean any algorithm in which the 
argument of the trigonometric functions in the velocity update (\ref{vb}) is not directly defined 
as $\ta=\w(\br)\dt$, but as some other functions of $\ta$.

\begin{figure}[hbt]
	\includegraphics[width=0.80\linewidth]{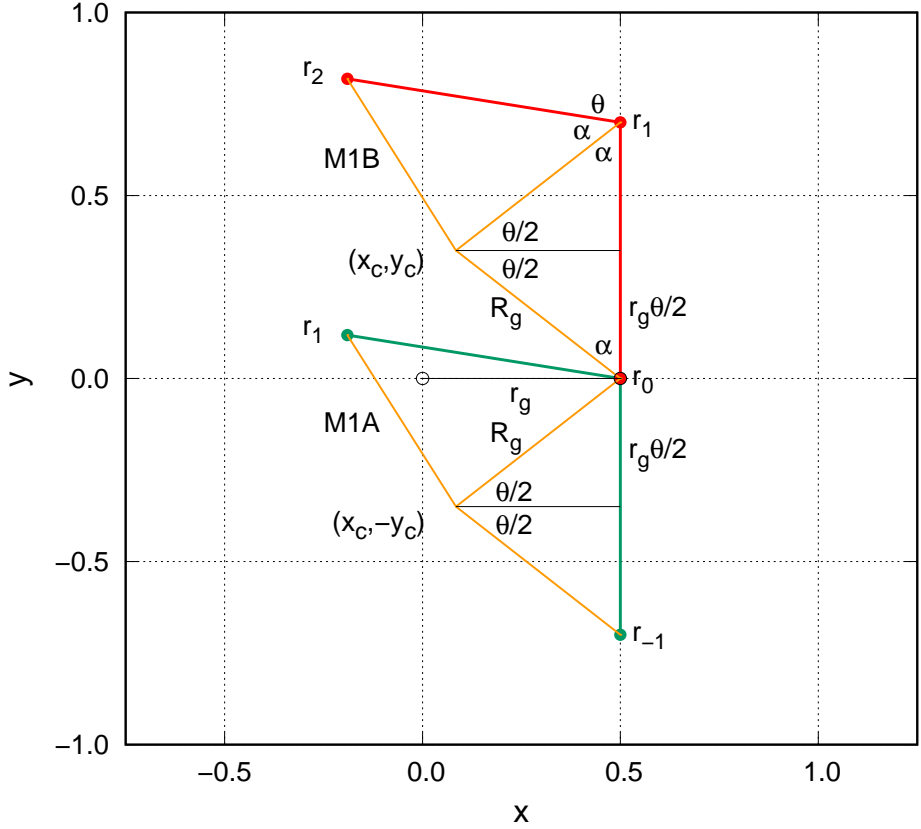}
	\caption{ (color online)
		The anatomy of first-order magnetic solvers M1A and M1B.	
	}
	\la{onedig}
\end{figure}

For clarity we will begin with magnetic field only integrators of the form (\ref{mag}). 
The two basic first-order approximations are
\be
\ct_{1A}=\e^{\dt T}\e^{\dt V_B}\quad{\rm and}\quad \ct_{1B}= \e^{\dt V_B}\e^{\dt T}
\ee
producing the following two sequential {\it magnetic} field integrators M1A,
\ba
\bv_1&=&\bv_B(\br_0,\bv_0,\dt)\nn\\
\br_1&=&\br_0+\dt \bv_1
\la{m1a}
\ea
and M1B,
\ba
\br_1&=&\br_0+\dt \bv_0\nn\\
\bv_1&=&\bv_B(\br_1,\bv_0,\dt).
\la{m1b}
\ea
Eq.(\ref{vb}) shows 
that the local magnetic field only rotates the perpendicular velocity component 
by $\ta=\w(\br)\dt$, leaving its parallel component and magnitude unchanged.
M1A first rotates $\bv_0$ by $\ta$ then moves to the new position along
the rotated velocity. M1B first moves to the new position using the present velocity, 
then rotates it after arrival.
In our naming scheme, the
suffix A or B denotes the algorithm whose first step is updating the velocity or the position,
respectively.

The working of these two algorithms can be easily analyzed for a negatively charged particle 
in a constant magnetic field in the $\bk$ direction, as shown in Fig.\ref{onedig}. 
When the particle is at $\br_0=(r_0,0)$, moving with tangential {\it vertical} velocity 
$\bv_0=(0,v_0)$ on the gyro-circle 
with radius $r_g=v_0/\w$, M1B would move it in time $\dt$, a vertical distance
$v_0\dt=r_g\ta$ to $\br_1$. At $\br_1$, it would rotate the velocity {\it from the vertical}
by $\ta$ and move it to $\br_2$.
Since both $\br_0$ and $\br_1$ must be on the algorithm's gyro-circle of radius $R_g$ centered 
at $(x_c,y_c)$, both must be equidistant from $(x_c,y_c)$. This means that $(x_c,y_c)$
must lie on the perpendicular bisector of $\br_1-\br_0$, and therefore 
\be
y_c= r_g\ta/2.
\la{ycb}
\ee
At $\br_1$, $\ta$ is the rotation angle from the vertical and the supplementary angle
to it is $2 \alpha$.
Since $\ta+2\alpha=\pi\rightarrow \ta/2+\alpha=\pi/2$,
the bisector's angle with either $R_g$ on its sides is $\ta/2$ and therefore 
$R_g\sin(\ta/2)=r_g\ta/2$, or
\be
R_g=r_g\frac{\ta/2}{\sin(\ta/2)}=r_g(1+\frac{\ta^2}{24}+\cdots).
\la{mrg}
\ee
This also means that the length of the bisector is $R_g\cos(\ta/2)$ and hence
\ba
x_c&=&r_g-R_g\cos(\ta/2)\nn\\
&=&r_g\left(1-\frac{\ta/2}{\tan(\ta/2)}\right)= r_g\left(\frac{\ta^2}{12}+\cdots\right).
\la{xc}
\ea 
(Note that since $\ta/2+\alpha=\pi/2$, if one were to rotate $\br_1-\br_0$ by $\ta/2$,
then $y_c=0$. Also, if one down shifts $y\rightarrow y-r_g\ta/2$ by starting out at the
midpoint of $\br_1-\br_0$, then also $y_c=0$. These two cases will be considered in Sect.\ref{sec}.)

\begin{figure}[hbt]
	\includegraphics[width=0.80\linewidth]{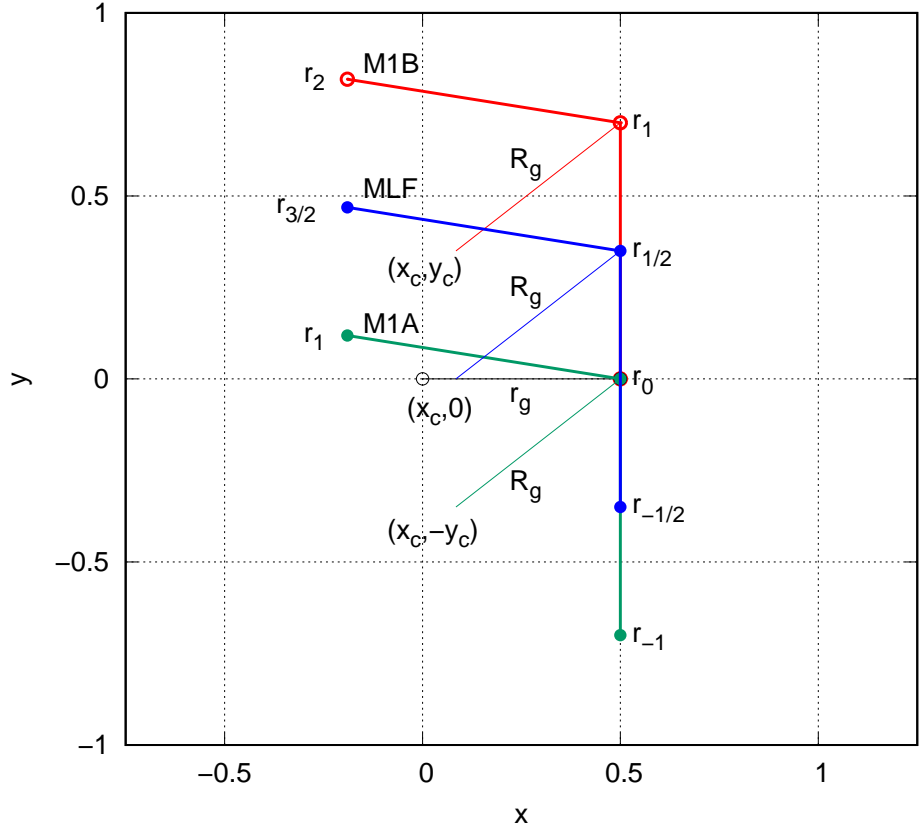}
	\caption{ (color online)
		The Magnetic Leap Frog algorithm MLF as compared to M1A and M1B.	
	}
	\la{lfdig}
\end{figure}

Similarly, for M1A, in order for the velocity to be rotated at $\br_0$, its previous 
position must be at $\br_{-1}$. It therefore follows that the y-coordinate of its 
gyro-center must be
\be
y_c=-r_g\ta/2
\la{yca}
\ee
but with the same $R_g$ and $x_c$ as given by (\ref{mrg}) and (\ref{xc}).
Since the exact cyclotron orbit must have $x_c=y_c=0$ and $R_g=r_g$,
(\ref{ycb})-(\ref{yca}) are the defining errors of these two basic
algorithms. As expected, the first-order (in $\dt$) errors 
$y_c=\pm r_g\ta/2$ are opposite in sign, while those of
$x_c$ and $R_g$ are of higher, even order in $\dt$. 

In additional to these two first-order algorithms, one also has the structurally 
similar leap frog algorithm defined on staggered time steps. Consider the case
where velocities are defined only on integer time-steps $n\dt$ and 
positions only at half-integer time-steps $(n-1/2)\dt$. A sequential algorithm would 
naturally be
\ba
\br_{n+1/2}&=&\br_{n-1/2}+\dt \bv_n\la{lfp}\\
\bv_{n+1}&=&\bv_B(\br_{n+1/2},\bv_n,\dt).
\la{mlf}
\ea
We will refer to this algorithm as MLF (magnetic leap frog).
MLF is structurally similar to M1B, except that its positions are symmetric about
the velocity. Its anatomy is compared to that of M1A and M1B in Fig.\ref{lfdig}.
Given $\br_0$ and $\bv_0$ for M1B, ones see that MLF corresponds to M1B
starting at a half time-step backward position $\br_{-1/2}=\br_0-(\dt/2)\bv_0$,
whereas M1A corresponds to M1B starting at a full time-step backward position $\br_{-1}=\br_0-\dt\bv_0$.
Since $\br_{-1/2}$ and $\br_{1/2}$ are symmetric about $\bv_0$, the
first order error $y_c$ vanishes for MLF. Thus MLF is a second order algorithm.
{\it However, despite MLF being second-order, it has the same errors  
$R_g$ (\ref{mrg}) and $x_c$ (\ref{xc}) as M1A and M1B.}

Conventionally, higher order methods would eliminate errors  
$R_g$ and $x_c$ order by order in $\ta$. However,
trajectories in a constant magnetic field have two distinct
motions: translation by $v_0\dt=r_g\ta$ and rotation by angle $\ta=\w\dt$.
It is only because $v_0$ remains unchanged in a constant magnetic field that
both motions are proportional to the same $\ta$. In principle, and in conventional dynamics, 
there is no such coupling between the two that would force the same $\ta$ on both. 
One therefore has this residual freedom of decoupling both motions to reduce the errors of 
$R_g$ and $x_c$ to all orders of $\ta$! {\it This is the foundational insight by which 
this work explains the existence of Boris solvers}, which is distinct from
conventional derivations based on {\it ad hoc} modifications of finite-difference
schemes.

First, one can decouple the rotation angle $\ta$
in trigonometric functions to an effective angle $\ta_B(\ta)$.
From (\ref{xc}), the choice of
\be
\tan(\ta_B/2)=\ta/2
\la{ban}
\ee
would force $x_c=0$. If $x_c=0$, then from Fig.\ref{onedig}, $R_g$ is the
hypotenuse of a right triangle with base $r_g$ and height $r_g\ta/2$:
\be
R_g=r_g\sqrt{1+\frac{\ta^2}4 }.
\la{brg}
\ee 
This choice (\ref{ban}) means that for (\ref{m1a}) and (\ref{m1b}), 
the rotation angle $\ta$ in (\ref{vb}) is to be replaced by 
the Boris angle $\ta_B$, yielding
\ba
\bv_B(\br,\bv,\dt)
&=&\bv_{||}+\cos\ta_B\bv_{\perp}+\sin\ta_B(\bh\times\bv_{\perp})
\la{brt}
\ea
where
\ba
\sin\ta_B&=&\frac{2\tan(\ta_B/2)}{1+\tan^2(\ta_B/2)}
= \frac{\ta}{1+\ta^2/4}\la{tsin}\\
\cos\ta_B&=&\frac{1-\tan^2(\ta_B/2)}{1+\tan^2(\ta_B/2)}
= \frac{1-\ta^2/4}{1+\ta^2/4}.
\la{tcos}
\ea
This rotation angle replacement $\ta\rightarrow\ta_B$ in M1A, M1B and MLF then
produces Boris solvers B1A, B1B and BLF. The last being the conventional LF Boris
solver. Each Boris solver is uniquely 
characterized by its error in a constant magnetic field. For B1A and B1B, they are errors
in $y_c$ and $R_g$. For BLF, its error is only $R_g$.

Second, from (\ref{mrg}) one can force $R_g=r_g$ by defining a new
angle $\ta_C$, such that
\ba
\sin(\ta_C/2)=\ta/2,\quad\cos(\ta_C/2)=\sqrt{1-\ta^2/4},\la{can}
\ea
and consequently,
\ba
\sin\ta_C=\ta\sqrt{1-\ta^2/4},\quad \cos\ta_C=(1-\ta^2/2).
\la{crot}
\ea
In this case,
\be
x_c=r_g(1-\cos(\ta/2))=r_g(1-\sqrt{1-(\ta/2)^2}),
\ee
which limits the algorithm to $|\ta|\le 2$.
The resulting three fundamental algorithms with rotation angle $\ta_C$
are previously unknown Boris type solvers and will be referred to as 
C1A, C1B and CLF. They are characterized by having errors in the
gyrocenter $(x_c,y_c)$ for  C1A and C1B, but only the $x_c$ error for CLF.

\begin{figure}[hbt]
	\includegraphics[width=0.80\linewidth]{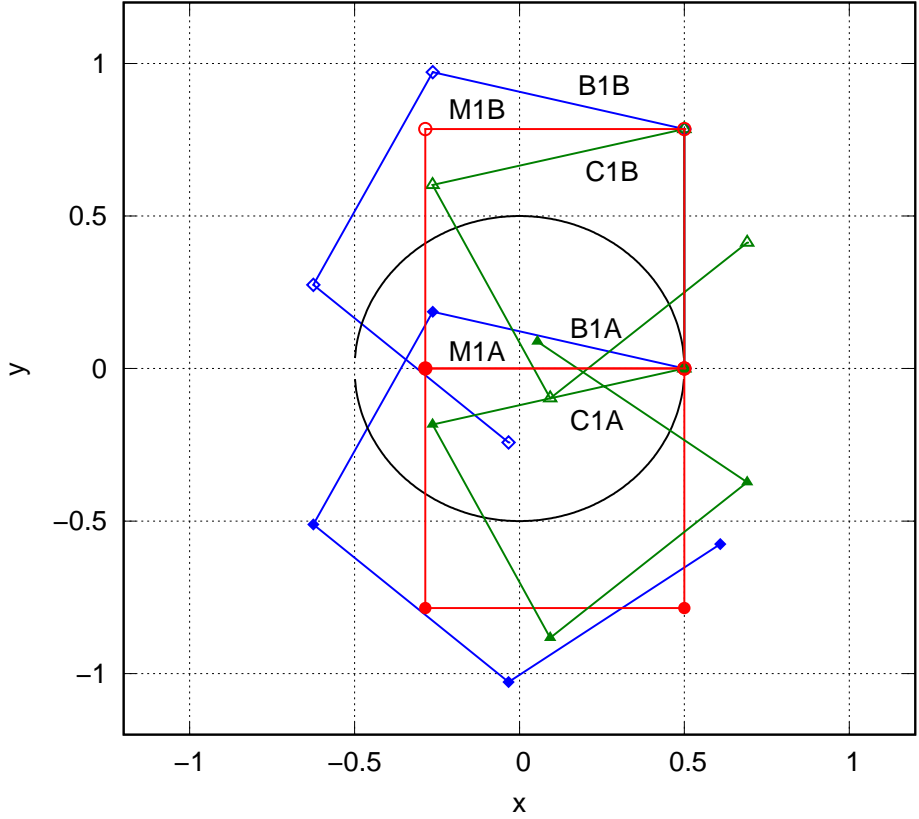}
	\caption{ (color online)
		The orbits of six first-order magnetic field algorithms at a large
		$\dt=\pi/4$.  		
	}
	\la{alg1}
\end{figure}

To see the working of these algorithms, consider the case of an electron 
in a constant magnetic field with $\w=2$, $\bh=\hat {\bf z}$, $\br=(x,y)$, $\bv=\bv_\perp=(v_x,v_y)$,
with initial velocity $\bv_0=(0,v_0)$, $\br_0=(r_g,0)$, where $v_0=1$ and where
$r_g=v_0/\w=1/2$ is the gyro-radius.
Take a large $\dt=\pi/4$, $\ta=\w\dt=\pi/2$ so that the trajectory of M type algorithms
would rotate through $90^\circ$ 4 times to complete one orbit.
For non-leapfrog algorithms, the M algorithms 
are the two (red) square orbits shown in Fig.\ref{alg1}.
The two blue and green orbits are those of B and C solvers.
All six algorithms obviously exhibit the gyrocenter error $y_c=\pm r_g\ta/2$.
The orbits of B and C solvers are tilted backward and forward as compare to 
the M algorithms because their decoupling angle
\ba
\ta_B&=&2\tan^{-1}(\ta/2)=\ta-\frac{\ta^3}{12}+\cdots\nn\\
\ta_C&=&2\sin^{-1}(\ta/2)=\ta+\frac{\ta^3}{24}+\cdots
\ea
lags or leads the correct angle. 

\begin{figure}[hbt]
	\includegraphics[width=0.80\linewidth]{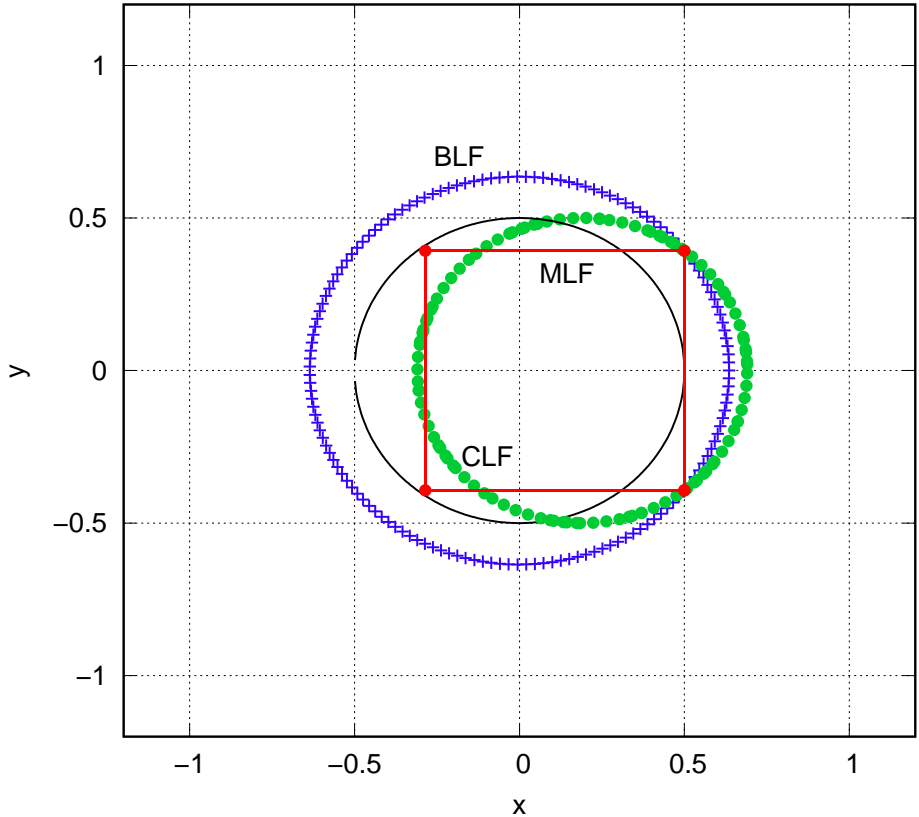}
	\caption{ (color online)
		The orbits of three leap frog type algorithms at a large
		$\dt=\pi/4$ iterated many more times to trace out their orbits.  		
	}
	\la{alglf}
\end{figure}

The three leap frog algorithms with more iterations are plotted in Fig.\ref{alglf}
without connecting lines (except for MLF).
Since MLF has the correct angle, it will just keep on tracing out a $y_c=0$ square.
This graph verifies that BLF has a centered, but large gyroradius $R_g$,
while CLF has the correct $r_g$ radius but is off-center to the right by $x_c$. 

Since these nine integrators only rotate the velocity vector, all are exact energy conserving for
a general magnetic field. In the limit of $\dt\rightarrow 0$, all nine algorithms will converge 
onto the exact gyro-orbit. 

In the limit of large $\dt$, the C algorithms are limited to $\dt\le 2/\w$, otherwise, $\ta_C$
cannot be defined by (\ref{can}). For the M algorithms, their gyro-radius (\ref{mrg}) can be
arbitrarily large near $\ta=n 2\pi$ and is always $\ge r_g\ta/2$. For the B algorithms,
their gyro-radii given by (\ref{brg}) grow linearly as $r_g\ta/2$ at large $\dt$.

\begin{figure}[hbt]
	\includegraphics[width=0.48\linewidth]{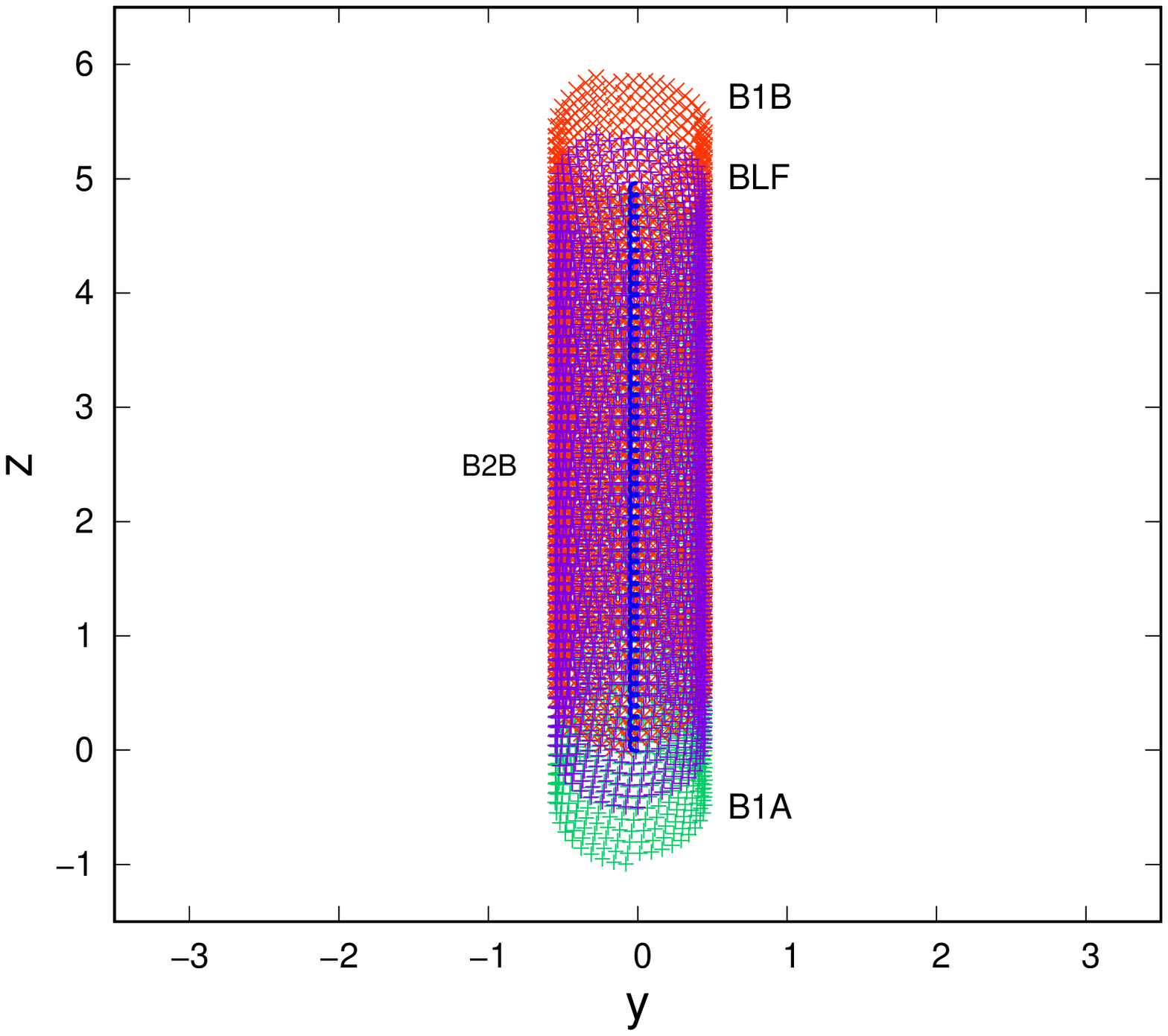}
	\includegraphics[width=0.48\linewidth]{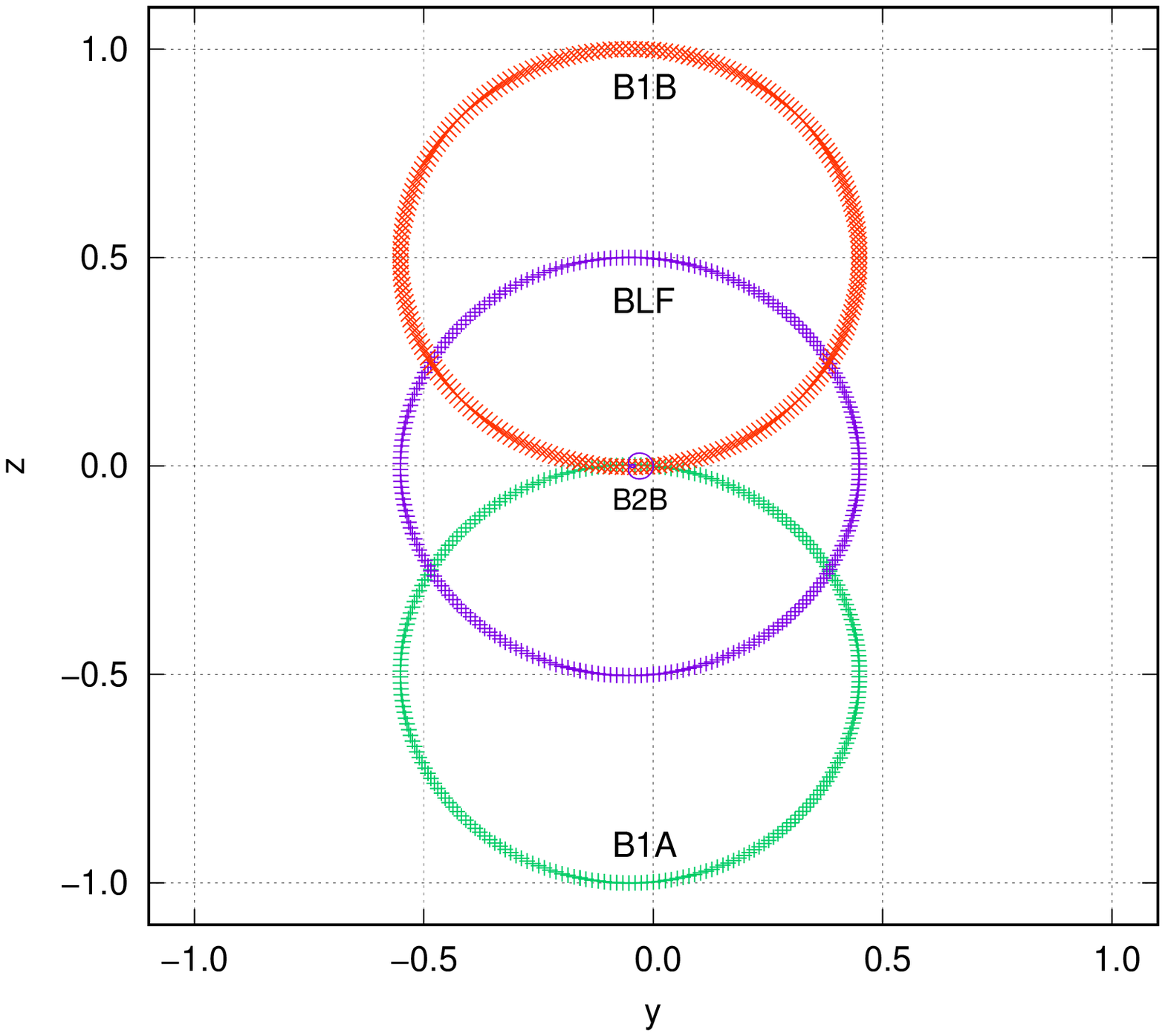}
	\caption{ (color online) {\bf Left} (a): Trajectories of
		solvers B1A, B1B, BLF and B2B at $\dt=0.5$, reproducing 
		Parker and Birdsall's\cite{par91} gradient B drift calculation Fig.3. 
		{\bf Right} (b): Collapsing the trajectories by removing the vertical drift in $z(t)$.
		The smaller font B2B indicates the tiny blue collapsed cycloid.
		See text for details. 		
	}
	\la{pfthree}
\end{figure}
 
To see which of the B algorithm has been historically regarded as ``the Boris solver",
we apply B1A, B1B, BLF and B2B (to be derived in the next section)  
to the case of a non-uniform magnetic field
{\bf B}$=(100-25 y)\hat {\bf x}$, taken from Parker and Birdsall's \cite{par91}
Fig.3, with $\br_0=(0,0,0)$ and $\bv_0=(0,0,2)$. For this case, near $y=0$,  
$\w=100$, $\dt=0.5$, $\ta=50$, $2 r_g=4/100=0.04$, $2R_g=1.0008$ and 
$y_c=\pm r_g\ta/2=\pm 0.5$. 
The resulting trajectories are as shown in Fig.\ref{pfthree}(a).
By comparing Fig.\ref{pfthree}(a) to Parker and Birdsall's \cite{par91} Fig.3, 
it is easy to see that Parker and Birdsall's Boris solver is B1B 
and {\it not} BLF. This is because Parker and Birdsall's trajectory clearly has
an off-set of 0.5 {\it above} the gyrocenter of the smaller $\dt$ trajectory,
exactly the same way as B1B's gyrocenter is above that of B2B in Fig.\ref{pfthree}(a). 
By removing the vertical drift of $0.004941 t$ (rather than $0.005 t$),
the trajectories collapse back onto themselves as shown in Fig.\ref{pfthree}(b). 
The gyroradius and the vertical gyrocenter off-set errors for B1A, B1B and BLF 
are all as predicted.
(The slight horizontal center off-set error of $\approx -0.05$ due to the nonuniform
magnetic field is not accounted for by the above error analysis.)

The collapsed trajectory of B2B is a cycloid, with a maximum horizontal separation
of exactly $2 r_g=0.04$, but a vertical diameter of $\approx 0.06$. Its gyroradius
is therefore much closer to the exact and $\approx 20$ times smaller than those 
of B1A, B1B and BLF in Fig.\ref{pfthree}(b). (It may not be visible unless the figure 
is greatly enlarged.) 

\begin{figure}[hbt]
	\includegraphics[width=0.49\linewidth]{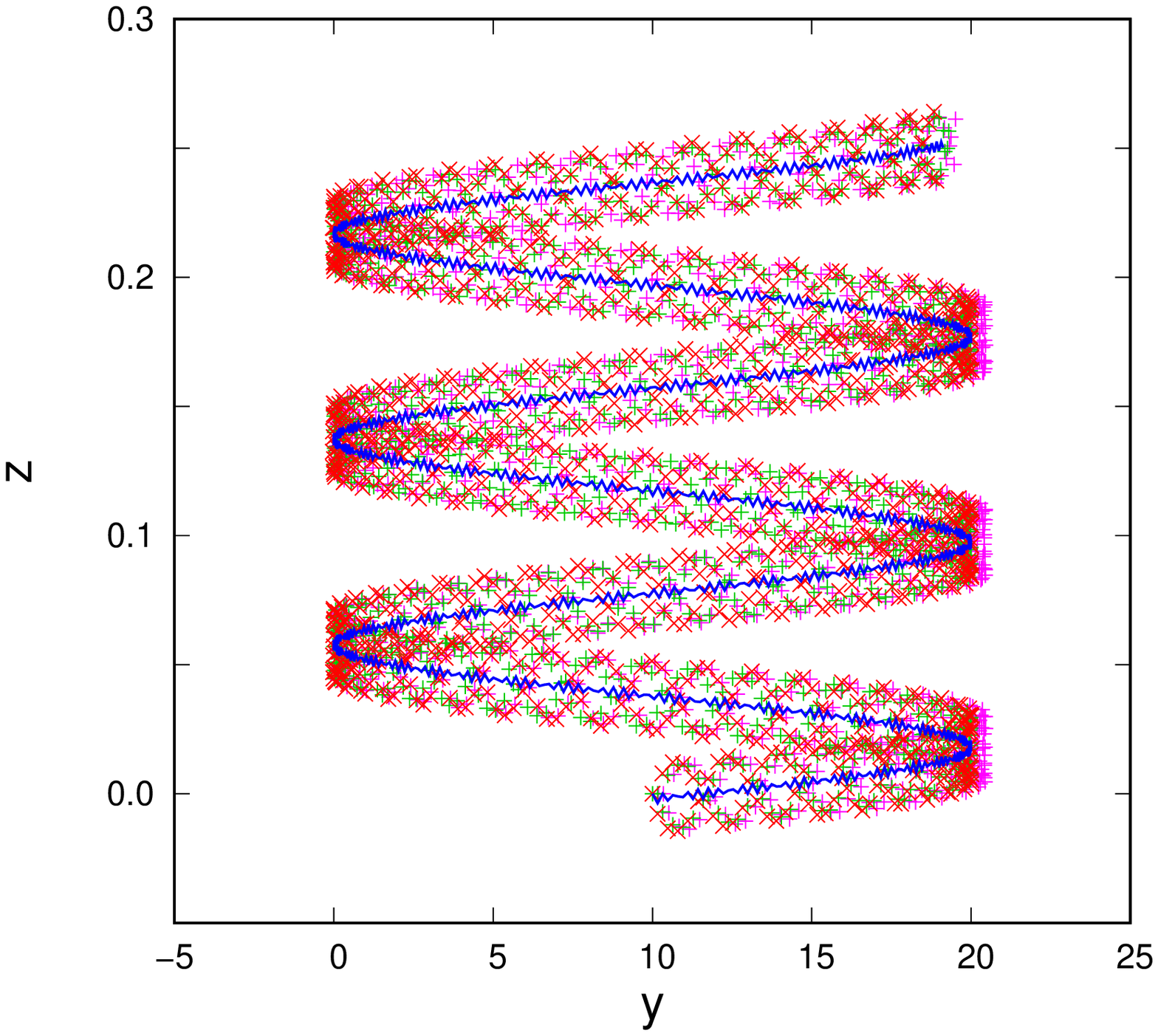}
	\includegraphics[width=0.49\linewidth]{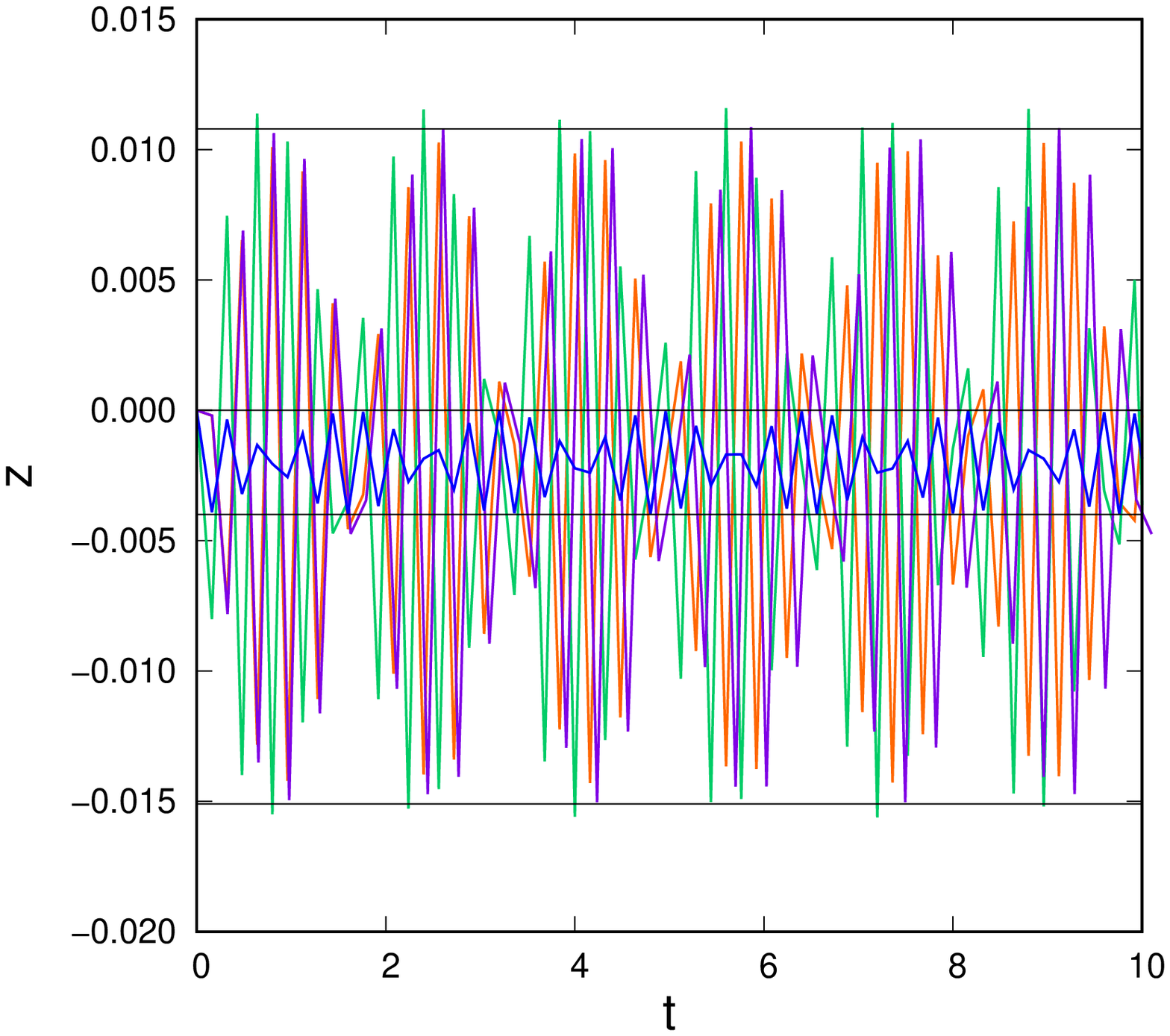}
	\caption{ (color online) {\bf Left} (a):
		Trajectories from B1A (green), B1B (red), BLF (purple) and B2B (blue)
		at $\dt=0.16$, reproducing Parker and Birdsall's\cite{par91}
		Fig.2. {\bf Right} (b): Oscillations of $z(t)$ after removing the vertical drift. 	
	}
	\la{pftwo}
\end{figure}

Next, we check all four solvers against Parker and Birdsall's\cite{par91}
curvature drift calculation due to a magnetic field 
${\bf B}(\br)= (800/r)\hat\bta$ from a line current flowing along $\hat {\bf z}$ 
at $\br=(10,10,0)$ with $\br_0=(0,10,0)$ and $\bv_0=(0.16,1,0)$. The time
step is nearly twice the gyroperiod at the starting position. 
The y-z trajectories of B1A, B1B and BLF
all overlap, with no discernible gyrocenter off-sets. 
B2B's oscillations are much, much smaller. The oscillation sizes can again be
determined by removing the vertical drift ($0.0012648\,t$) in Fig.\ref{pftwo}(b).
For $z$ along $\br=(10,10,z)$, $\w=80$, $v_{\perp}=0.16$, 
the gyro-radius is $0.16/80=0.002$. As shown in Fig.\ref{pftwo}(b),
B2B's oscillation is precisely within the gyro-diameter of $2r_g=0.004$. 
At $\dt=0.16$, $\ta=\w\dt=12.8$, the first-order gyro-diameter is
$2R_g=0.0259$. The top and bottom lines 
in Fig.\ref{pftwo}(b) are at $z=0.0108$ and $z=-0.0151$ respectively,
marking the diameter of BLF as $0.0259$! (This graphical fitting was done
prior to knowing the values of $2R_g$.) Similarly, B1A's and B1B's diameters
were graphically fitted to be $0.0272$ and $0.0247$ respectively, whose average,
surprisingly, is also $\approx 0.0259$. (This is also similar to the case of
the polarization drift in Sect.\ref{seceb}.)
Again, B2B tracks the correct local gyrocircle at large values of 
$\dt$, an order of magnitude better than B1A, B1B and BLF.

In the above two historical calculations, despite being second-order, BLF's trajectory, 
unlike that of B2B, is not any better than those of B1A and B1B. This is not true in general.
Consider the case of planar motions in a Gaussian magnetic field,
\be
\B(\br)=B_0\e^{-r^2}\hat{\bf z}.
\ee 
Let $\br_0=(r_g,0)$ and $\bv_0=(0,v_0)$. Since the magnetic field is radially symmetric,
the magnetic field is the same all along the circumference of the gyrocircle. Therefore,
$r_g$ is the same as that of a constant magnetic field of magnitude $B_0\e^{-r_g^2}$:
\be
r_g=\frac{v_0}{\w}=\frac{v_0}{B_0}\e^{r_g^2}.
\ee
Thus, given any $r_g$, the required orbital velocity is
\be
v_0={B_0}r_g\e^{-r_g^2},
\ee
with period $T=2\pi r_g/v_0$. Choose $r_g=0.1$, then from (\ref{brg}) one can choose
$\dt=(\sqrt{99}/\pi)T$ so that $R_g=1$. The resulting trajectories of all four Boris solvers are
shown in Fig.\ref{gauss}. B2B is on the correct orbit with radius $r_g$. BLF is on the wrong orbit
with radius $R_g$.
However, in additional of having the wrong radius $R_g$, B1A and B1B also have the 
off-set errors 
$y_c=\pm r_g\sqrt{99}\approx \pm 1$. Because of this off-set error, their trajectories are also rotated 
by the gradient B drift. BLF suffered no such rotation.

\begin{figure}[hbt]
	\includegraphics[width=0.80\linewidth]{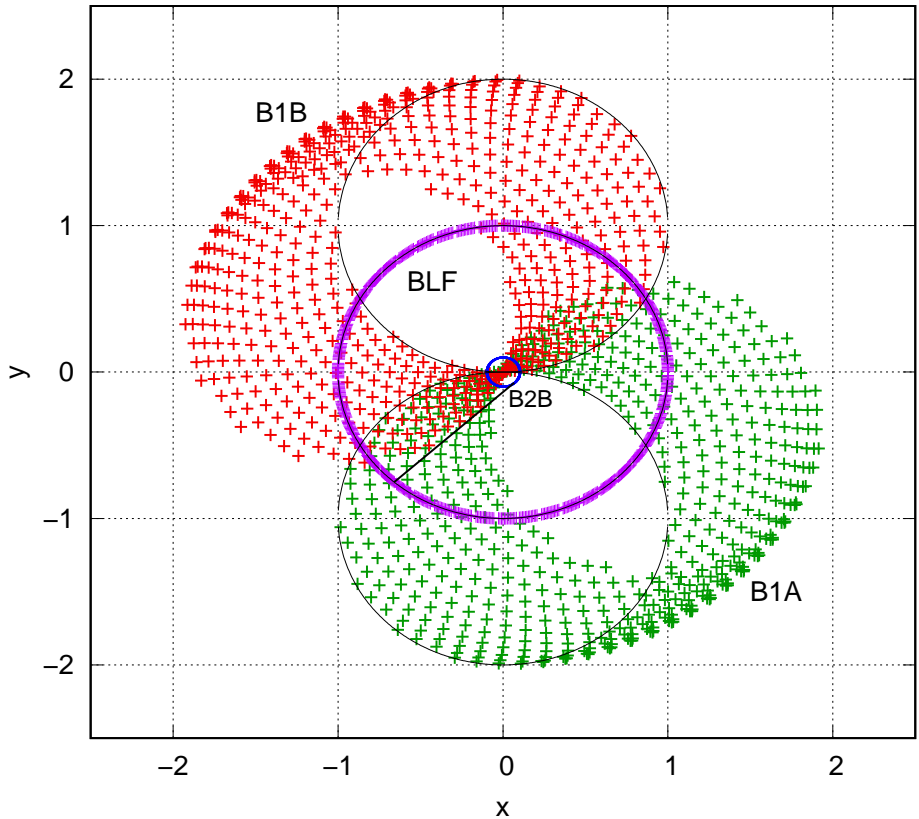}
	\caption{ (color online)
		Trajectories of all four Boris solvers in a radial Gaussian magnetic field.
		The line connecting BLF and B2B will be explained in Sect.\ref{diff}.
	}
	\la{gauss}
\end{figure}

This large gyroradius $R_g$, given by (\ref{brg}), is cited as that of 
the ``Boris solver'' in Refs.\onlinecite{bir85,par91,vu95,vay08,qin13,zen18,ric20}. 
There is no such large gyroradius error in 
the second-order solver B2B. The difference between BLF and B2B will
be explained in Sect.\ref{diff}.

\section {Second-order magnetic field integrators}
\la{sec}

The leap frog construction can eliminate first-order errors $y_c=\pm r_g\ta/2$ 
by use of staggered time steps (\ref{mlf}), as shown in Fig.\ref{alglf}.
However, as also shown in Fig.\ref{alglf}, even the adoption of either Boris angle
cannot completely get rid of both errors $R_g$ and $x_c$.
Here, we show how this can be done by symmetric second-order methods
which are superior to the use of staggered time steps.

\begin{figure}[hbt]
	\includegraphics[width=0.80\linewidth]{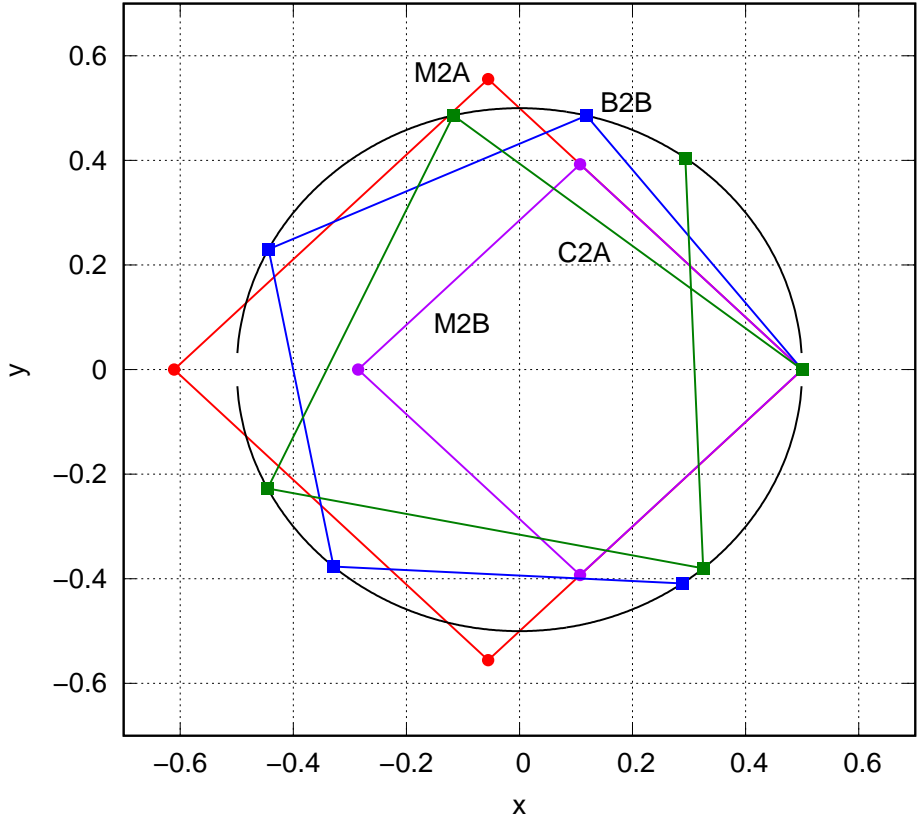}		
	\caption{ (color online)
		For the same configuration as Fig.\ref{alg1}, the
		two second-order algorithms M2A and M2B produce two square orbits
		having the correct rotation angles. C2A and B2B are the two second-order Boris solvers with 
		trajectories exactly on the gyro-circle but with out-of-phase rotation angles. 
	}
	\la{mb2}
\end{figure}
 
In sequential symplectic integrators\cite{yos93,chin20}, it is well known that 
first-order errors can be automatically removed by a time-symmetric 
concatenation of the two first-order methods, 
\be
\ct_{2A}=\e^{(\dt/2) V_B}\e^{\dt T}\e^{(\dt/2) V_B}
\quad{\rm and}\quad \ct_{2B}=\e^{(\dt/2) T}\e^{\dt V_B}\e^{(\dt/2) T}
\la{msec}
\ee
yielding the following second-order integrators M2A, 
\ba
\bv_{1/2}&=&\bv_B(\br_0,\bv_0,\dt/2)\nn\\
\br_1&=&\br_0+\dt \bv_{1/2}\nn\\
\bv_1&=&\bv_B(\br_1,\bv_{1/2},\dt/2)
\la{m2a}
\ea
and M2B,
\ba
\br_{1/2}&=&\br_0+\frac12\dt \bv_0\nn\\
\bv_1&=&\bv_B(\br_{1/2},\bv_0,\dt)\nn\\
\br_1&=&\br_{1/2}+\frac12 \dt \bv_1.
\la{m2b}
\ea
In order to facilitate comparison with staggered time-step algorithms,
sequentially updated variables in the above two algorithms have been subscripted by 
the accumulated time step of that variable.
For the same test problem as in Fig.\ref{alg1}, they now produce the two
upright square orbits as labeled in Fig.\ref{mb2}.
The glaring off-set errors $y_c=\pm r_g\ta/2$ in Fig.\ref{alg1} are now absent.

\begin{figure}[hbt]
	\includegraphics[width=0.48\linewidth]{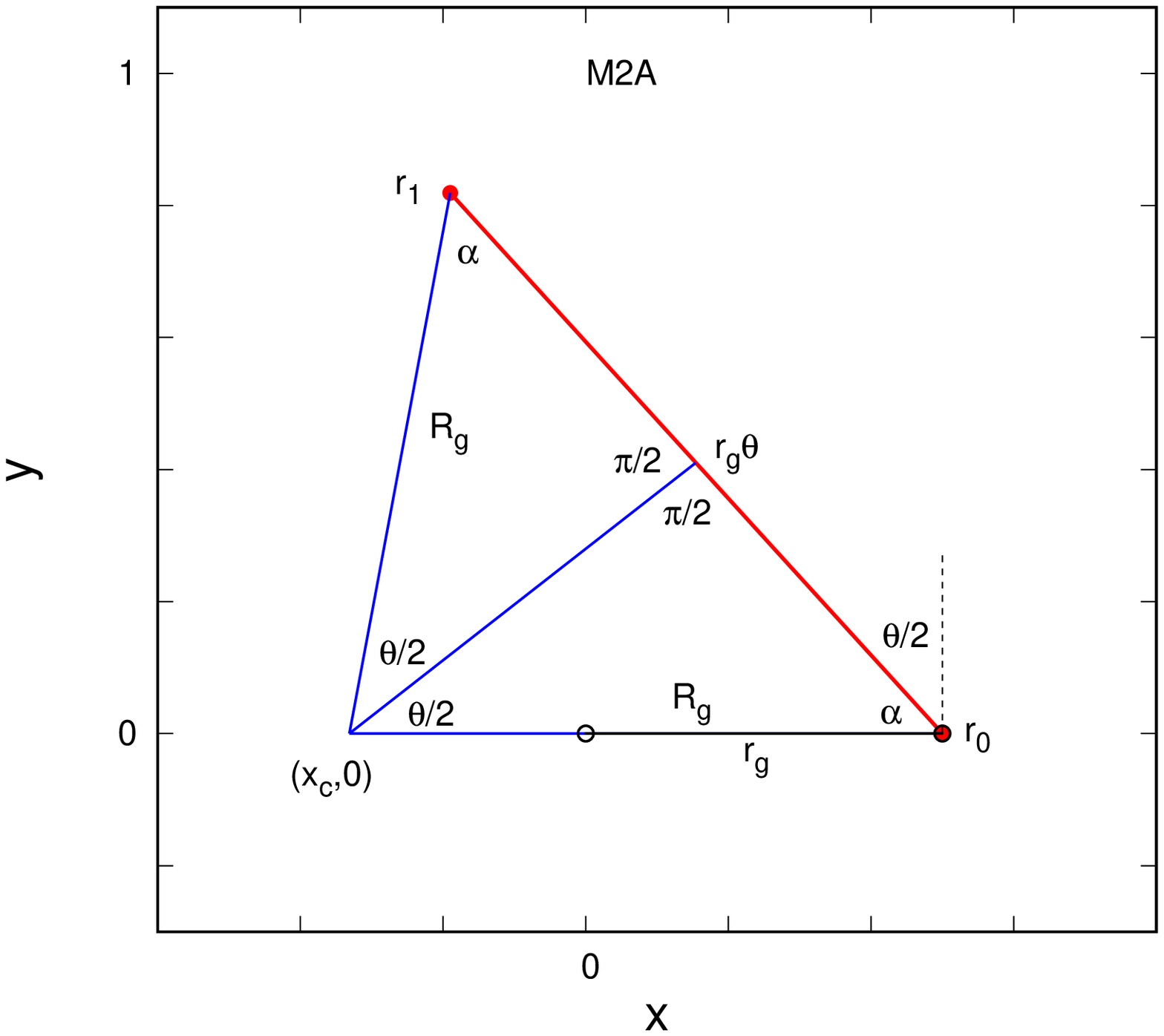}
	\includegraphics[width=0.48\linewidth]{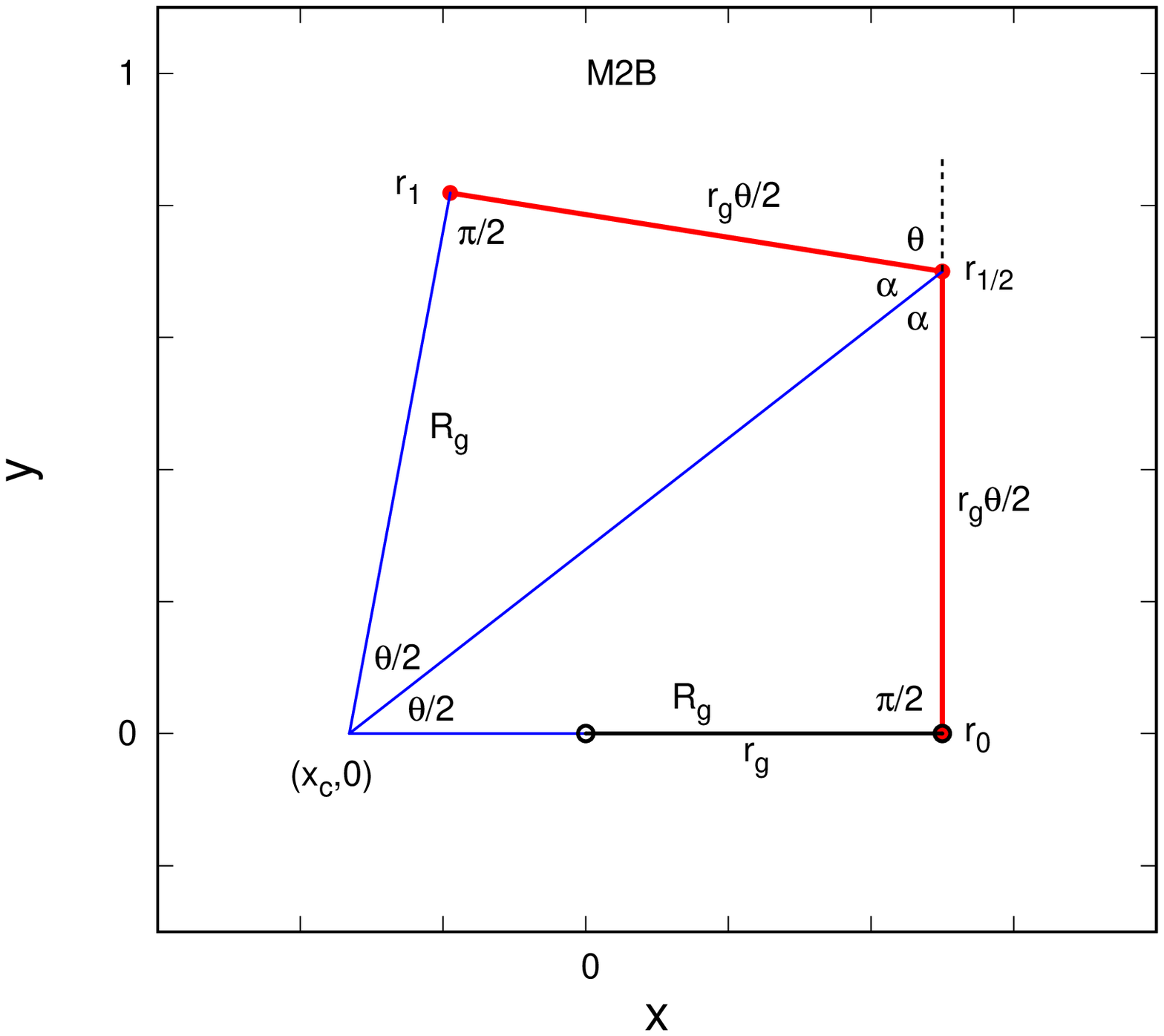}
	\caption{ (color online)
		The anatomy of second-order Poisson solvers M2A and M2B.	
	}
	\la{twodig}
\end{figure}

The anatomy of these two algorithms are shown in Fig.\ref{twodig}.
For M2A, the particle starts at $\br_0$, rotates by $\ta/2$, then travels the
full distance $r_g\ta$ to $\br_1$. This is just $\br_1-\br_0$ of M1B rotated by $\ta/2$
and therefore $y_c=0$. One then has the bottom right triangle with
\be
\sin(\ta/2)=\frac{r_g\ta/2}{R_g}\quad\rightarrow\quad R_g=r_g\frac{\ta/2}{\sin(\ta/2)}
\ee
and again
\be
x_c=r_g-R_g=r_g(1-\frac{\ta/2}{\sin(\ta/2)}).
\la{xca}
\ee
At first sight, M2A is no better than MLF, since it still has errors $R_g$ and $x_c$.
However, since the $y_c$ error in this case is eliminated by a rotation so that $R_g$ is 
along the $x$-axis, the $x_c$ error above has the same error dependence as $R_g$ 
(in contrast to (\ref{xc})),
and {\it both} can be simultaneously set to zero by 
the alternative Boris angle $\sin(\ta_C/2)=\ta/2$ !
This means that if $\br_0$ is initially at the gyrocircle, then
$\br_1$, and all subsequent positions must also be on the gyrocircle, as long as
$|\ta/2|\le 1$. We will refer to this algorithm as C2A. Note that for C2A, 
only its defining angle (\ref{can}) is needed in (\ref{m2a}), not its double angle (\ref{crot}).

Similarly, for M2B, from Fig.\ref{twodig}, the particle starts at $\br_0$,
moves along $\bv_0$ a distance $r_g\ta/2$ to $\br_{1/2}$, 
rotates $\bv_0$ by $\ta$, then travels $r_g\ta/2$ again to $\br_1$.
The points $\br_0$ and $\br_1$ here are just midpoints of M1B with downward shifted $y\rightarrow y-r_g\ta/2$,
resulting in $y_c=0$. 
From the base triangle, one now has
\be
\tan(\ta/2)=r_g(\ta/2)/R_g\quad\rightarrow\quad R_g=r_g\frac{\ta/2}{\tan(\ta/2)}
\ee
and
\be
x_c=r_g-R_g=r_g(1-\frac{\ta/2}{\tan(\ta/2)}).
\ee
Again, because $R_g$ is now along the $x$-axis, $x_c$ above has the same error dependence as $R_g$.
The original Boris angle $\tan(\ta_B/2)=\ta/2$ then also simultaneously eliminate both, but in this
case trajectories will be 
exactly on the gyrocircle for {\it all} $\dt$. We will refer to this algorithm as B2B,
or the {\it symmetric} second-order Boris solver.
The trajectories of C2A and B2B are as shown in Fig.\ref{mb2}. The rotation angles 
of M2A and M2B are again exactly correct, while those of C2A and B2B are ahead and behind
by approximately the same amount.

Algorithms B2A and C2B, corresponding to choosing the wrong Boris angle for M2A and M2B
will not yield trajectories on the gyro-circle. 
They are just phase-shifted versions of M2A and M2B and therefore
not shown in Fig.\ref{mb2}. All six algorithms will converge as second-order integrators
at small $\dt$. However, one perennial problem of 
plasma physics simulations is that one would like to use time steps not limited by the 
rapid local cyclotron motion and short gyro-period. The Boris solver B2B is unique
in that in the limit of $\ta\rightarrow\infty$,
$\ta_B\rightarrow\pi$, B2B's trajectory will just bounce back and forth nearly 
as straight lines across the diameter of the gyro-circle. Thus in contrast to
all other algorithms, only B2B's trajectory remains bounded 
to the exact orbit even as $\dt\rightarrow\infty$. 
As shown in Figs.\ref{pfthree}
and \ref{pftwo}, its gyro-radius remains nearly identical to the exact result 
even for a non-uniform magnetic field and is orders of magnitude smaller 
than those of B1A, B1B or BLF.

In a recent work, one of us has given an alternative derivation\cite{chin20b} of C2A and B2B
by requiring M2A's and M2B's trajectory to be exactly on the gyro-circle. 
That then automatically forces the gyro-center to the origin and $R_g=r_g$.
That derivation did not explain why one has to start with M2A and M2B.
The present derivation shows that
first-order algorithms have off-center errors $x_c$ and $y_c$. The error
$y_c$ must first be eliminated by symmetric second-order solvers M2A and M2B.
The errors $x_c$ and $R_g$ can then be eliminated simultaneously 
by a suitable choice of Boris angles, resulting in on-orbit trajectories.
The two derivations are therefore complementary.
A third derivation of C2A and B2B has been implicitly given in
Ref.\onlinecite{chin08} sometime ago. For completeness, that derivation will now be 
summarized in Appendix A.  

Algorithm B2B is cited as the Boris solver in Refs.\onlinecite{sto02,he15,kna15,rip18}.
Most think that B2B is just a reformulation\cite{kna15} of, or is ``essentially the same''\cite{rip18}
as, BLF. As shown in this work, this is not the case. BLF has the error gyroradius $R_g$ (\ref{brg})
while B2B does not. The difference between the two will be explained in the next section.  

The fact that B2B trajectory in a constant magnetic field is exactly on the gyrocircle
seemed not to be widely known after Boris superseded Buneman's derivation\cite{bir85}, 
otherwise, it would not have been necessary for Stoltz, Cary, Penn and Wurtele\cite{sto02} 
to explicitly verify that again in 2002. This on-orbit property is also {\it not} noted in some 
recent publications\cite{zen18,rip18,ric20}. This may also be due to the fact that many authors
were not aware of the difference between BLF and B2B.

\section{Leap frog Boris and symmetric Boris are different algorithms}
\la{diff} 

As shown in Sect.\ref{first}, the leap frog Boris solver BLF, as originally formulated by Boris\cite{bor70},
and widely disseminated by Birdsall and Langdon\cite{bir85}, has the large $R_g$ error and is not on-orbit.
Yet, many publications\cite{sto02,he15,kna15,rip18} that use the on-orbit solver B2B, do not
distinguish the latter as being different from the original ``Boris solver''. In this section, we make it
absolutely clear that the two are different algorithms having different gyroradii.

Consider iterating B2B in its operator form in a constant magnetic field
\be
\ct_{2B}^n=\cdots |\e^{(\dt/2) T}\e^{\dt V_B}\e^{(\dt/2) T}|\e^{(\dt/2) T}\e^{\dt V_B}\e^{(\dt/2) T}
|\e^{(\dt/2) T}\e^{\dt V_B}\e^{(\dt/2) T},
\la{al2b}
\ee
where each vertical bar $|$ indicates the end point of each iteration where $\br_n$ and $\bv_n$
are outputted at integer time steps beginning with $n=1$. The rotating angle in $\e^{\dt V_B}$
is $\ta_B$. 

Iterating the leap frog solver BLF corresponds to iterating B1B with an initial
half time-step backward position:
\ba
\ct_{LF}^n&=&\cdots|\e^{\dt V_B}\e^{\dt T}|\e^{\dt V_B}\e^{\dt T}|\e^{\dt V_B}\e^{\dt T}\e^{-(\dt/2) T},\nn\\
&=&\cdots|\e^{\dt V_B}\e^{\dt T}|\e^{\dt V_B}\e^{\dt T}|\e^{\dt V_B}\e^{(\dt/2) T}.
\la{allf}
\ea
At every end point, because of the the initial $\e^{-(\dt/2) T}$, the position variable is always at 
half integer time steps $\br_{n-1/2}$, while $\bv_n$ remains at integer time steps. 
Because of this, positions at integer time step
$\br_{n}$ {\it do not exist} for BLF.
Any attempt to define an integer time-step position $\br_{n}$ for BLF, is an {\it ad hoc}
alteration of the algorithm, making it no longer a leap frog algorithm.
For example, one can define the non-existent $\br_n$ in BLF as
\be
\br_n=\br_{n-1/2}+\frac12\dt\bv_n
\la{laststep}
\quad
{\rm and}\quad 
\br_{n+1/2}=\br_{n}+\frac12\dt\bv_n,
\ee
so that (\ref{lfp}) is satisfied. Introducing $\br_n$ this way is tantamount to splitting $\e^{\dt T}$
in (\ref{allf}) into two halves,
\ba
\ct_{LF}^n
&=&\cdots|\e^{\dt V_B}\e^{\frac12 \dt T}\e^{\frac12 \dt T}|\e^{\dt V_B}
\e^{\frac12 \dt T}\e^{\frac12 \dt T}|\e^{\dt V_B}\e^{(\dt/2) T}
\ea
and moving the end points to 
\ba
\ct_{LF}^n=
\cdots \e^{\dt V_B}\e^{\frac12 \dt T}|\e^{(\dt/2) T}\e^{\dt V_B}\e^{(\dt/2) T}|\e^{(\dt/2) T}\e^{\dt V_B}\e^{(\dt/2) T}
\ea
so that it now resembles (\ref{al2b}).
This last step, of moving the end points to the middle of $\e^{\dt T}$, where no such end point existed
in the original leap frog algorithm, fundamentally changed BLF to that of B2B. This is {\it not} a proof that
BLF is ``equivalent'' to B2B, but is an {\it ad hoc} derivation of B2B from BLF in the absence of 
a systematic formalism. 

\begin{figure}[hbt]
	\includegraphics[width=0.60\linewidth]{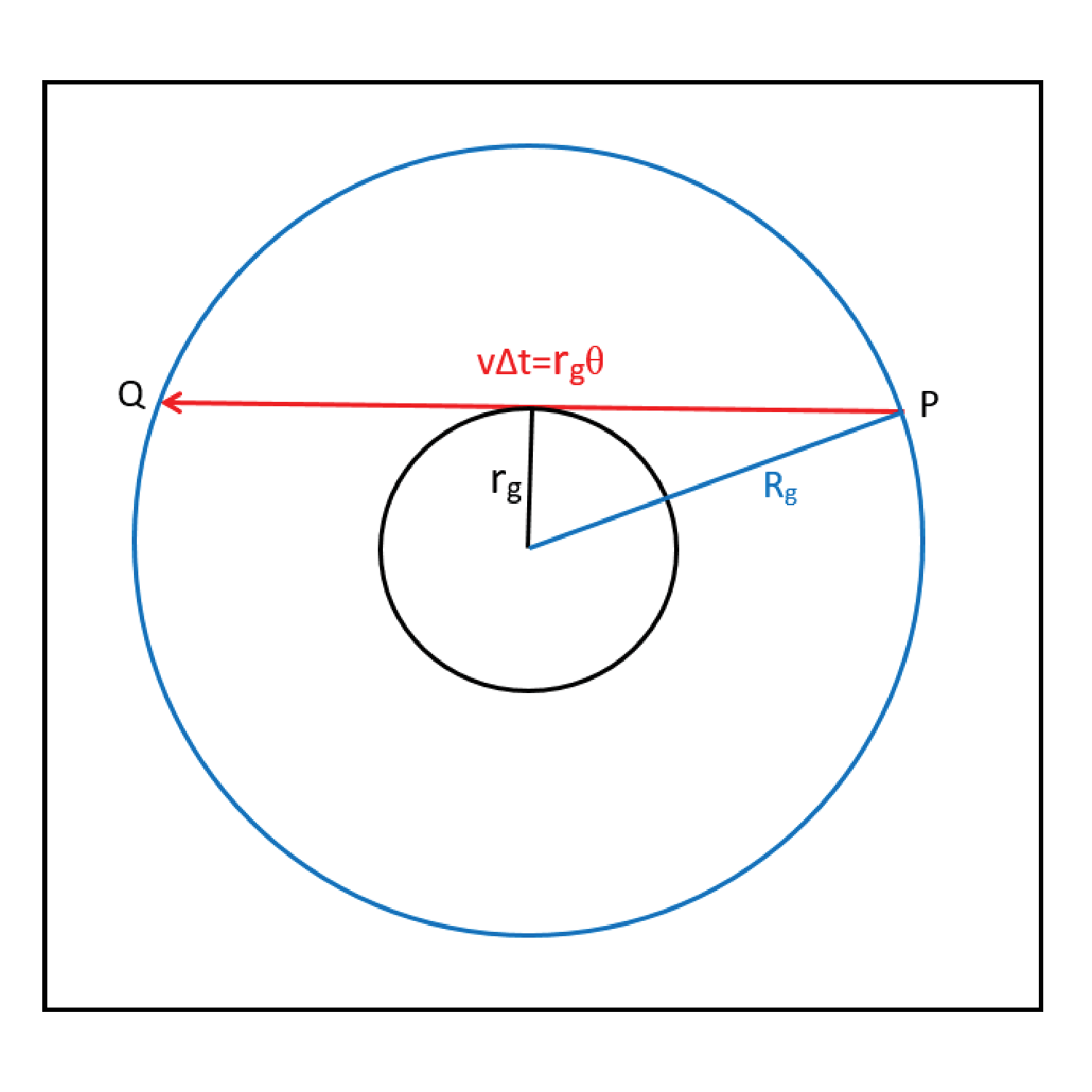}
	\caption{ (color online)
		The position P of BLF goes to Q along the chord of the $R_g$ circle after one $\dt$-step update.
		Each of B2B's position is only a half $\dt$-step to the mid-chord, resulting in the correct gyroradius $r_g$.		
	}
	\la{blfcomp}
\end{figure}

The reason why the two solvers have different gyroradii is extremely simple.
For B2B, if one outputs the position only at
the end of $N$ iterations, then (\ref{al2b}) is effectively
\be
\ct_{2B}^n=|\e^{(\dt/2) T}\e^{\dt V_B}\e^{\dt T}\cdots\e^{\dt V_B}\e^{\dt T}
\e^{\dt V_B}\e^{(\dt/2) T},
\la{blfend}
\ee
which is exactly like BLF of (\ref{allf}), {\it except} 
for the last position update $\e^{(\dt/2) T}$. Before this update, all positions and velocities of
B2B (\ref{blfend}) are {\it identical} to those of BLF (\ref{allf}). Both are on the same
gyrocircle of radius $R_g$ (\ref{brg}). 
For BLF, the next position, due to the next update $\e^{\dt T}$, will be the position Q,
at a distance $v\dt=(v/\w)(\w\dt)=r_g\ta$ from the original position P,
along a chord of the circle, as shown in Fig.\ref{blfcomp}. 
However, the position output by B2B, due to the final $\e^{(\dt/2) T}$, is only
a  distance $v\dt/2=r_g\ta/2$ to the middle of the chord, giving its distance from the center as
$\sqrt{R_g^2-r_g^2\ta^2/4}=r_g$! Every iterated position of BLF is always on the larger
$R_g$ circle. Every iterated position of B2B is that of BLF {\it plus} a half-time step position
to the mid-chord of the $R_g$ circle. Therefore, each position of B2B is {\it always} at the 
smaller $r_g$ circle. 

This is also illustrated in Fig.\ref{gauss}. After 600 iterations of BLF,
applying the final half-time step position update immediately drops the trajectory back to the correct
gyrocircle of B2B, as indicated by the black connecting line.    

\section {Second-order electric and magnetic field integrators}
\la{seceb}

For a combined electric and magnetic field, 
second order algorithms from (\ref{emag}) are 
\be
\ct_{2A}=\e^{(\dt/2) V_{BF}}\e^{\dt T}\e^{(\dt/2) V_{BF}}
\quad{\rm and}\quad \ct_{2B}=\e^{(\dt/2) T}\e^{\dt V_{BF}}\e^{(\dt/2) T},
\ee
which will be named as EM2A and EM2B.
Since the action of $\e^{\dt V_{BF}}$ on $\br$ and $\bv$ are known via (\ref{vbfeq}), the
algorithms are straightforwardly defined. 
However, we will give here a more intuitive derivation 
of (\ref{vap}) to make clear its connection with 
the original works of Buneman\cite{bun67} and Boris\cite{bor70}.

To minimize distractions, we will ignore the trivial motions parallel to the magnetic field
and assume that both $\bv$ and $\bac$ are perpendicular to $\B$.
To solve (\ref{em}), one can set 
\be
\bv=\tilde\bv+\bu\quad{\rm with}\quad\bu=\frac1{\w}\bh\times\bac
\ee
where $\bu$ is the ($\E\times\B$) drift so that (\ref{em}) is reduced to
\be
\frac{d\tilde\bv}{dt}=\w\bh\times\tilde\bv,
\la{vtil}
\ee
with a pure rotation solution
\be
\tilde\bv_1 =\cos\theta\tilde\bv_0+\sin\theta(\bh\times\tilde\bv_0)\equiv R(\br,\ta)[\tilde\bv_0]
\ee
where we have denoted the action of the rotation operator by a square bracket and that $\tilde\bv_0=\tilde\bv(0)$ and $\tilde\bv_1=\tilde\bv(\dt)$. This is then Buneman's\cite{bun67}
drift-subtracting velocity update:
\be
\bv_1-\bu =R(\br,\ta)[\bv_0-\bu].
\la{bun}
\ee
Here, we go beyond Buneman by letting the rotation operator acts on each velocity,
\ba
\bv_1&=&R(\br,\ta)[\bv_0-\bu]+\bu\la{st2p}\\
&=&R(\br,\ta)[\bv_0]+\bu-\cos\ta\bu-\sin\ta(\bh\times\bu)\la{st2q}\\
&=&R(\br,\ta)[\bv_0]+\frac1{\w}\left[(1-\cos\ta)\bh\times\bac+\sin\ta\bac\right],
\la{st3}
\ea
which is then just (\ref{vbfeq}) without the parallel motion.

Buneman's velocity update (\ref{bun}) was considered undesirable because $|\bu|\propto 1/\w\propto 1/B(\br)$
and is singular where $B(\br)\approx 0$. However, there is {\it no} such singularity in
(\ref{st3}), {\it after} $\bu$ has been rotated and combined. In the usual case of $\dt<1$, when 
$\w\propto B(\br)\rightarrow 0$, one also has $\ta=\w \dt\rightarrow 0$ and 
\be
\frac1{\w}\left[(1-\cos\ta)\bh\times\bac+\sin\ta\bac\right]\rightarrow
\dt\left[\ta(\frac12-\frac{\ta^2}{4!}+\cdots)\bh\times\bac+(1-\frac{\ta^2}{3!}+\cdots)\bac\right]
\la{smalldt}
\ee
with no singular terms. The only problem is when $\dt>>1$ such that when
$\w\rightarrow0$, $\ta$ remains finite.
 
Boris was widely credited for proposing the E-B splitting\cite{bor70,bir85} to
avoid Buneman's $1/\w$ singularity. However, there is {\it no} such singularity
in (\ref{st3}), {\it even if} $\dt>>1$, when the rotating angle is $\ta_B$! 
The splitting is completely unnecessary. 
Replacing the rotating angle in the drift term of (\ref{st3}) by $\ta_B$ gives,
without any approximation, the non-singular result  
\ba
\frac1{\w}\left[(1-\cos\ta_B)\bh\times\bac+\sin\ta_B\bac\right]
&=&\frac1{\w}\left(\frac{\ta^2/2}{1+\ta^2/4}\bh\times\bac+\frac{\ta}{1+\ta^2/4}\bac\right)\nn\\
&=&\dt\left(\frac{\ta/2}{1+\ta^2/4}\bh\times\bac+\frac{1}{1+\ta^2/4}\bac\right).
\la{nosp}
\ea
Buneman also used the Boris angle $\ta_B$ in (\ref{bun}) for ``cycloid fitting'',  
making the trajectory on-orbit for a constant $\E$ and $\B$ field. However, it was difficult to
see the cancellation of $\w$ without rotating and combining the drift term as in (\ref{st2q}).

This non-singular result (\ref{nosp}) can also be directly derived from Boris' original equation.
Boris'\cite{bor70} Eq.(22), corresponding to the velocity update
\be
\bv_1-\bv_0=\frac12 \ta\hat\B\times(\bv_1+\bv_0)+\dt\bac,
\ee
can be solved as a matrix equation
\ba
(1-\frac12\ta\C)\bv_1&=&(1+\frac12 \ta\C)\bv_0+\dt\bac,\nn\\
\bv_1&=&\frac1{(1-\frac12\ta\C)}[(1+\frac12 \ta\C)\bv_0+\dt\bac], 
\la{borrot}
\ea
when $\C$, the cross-product operator defined in Sect.\ref{om}, is regarded as a $3\times 3$ matrix. 
Boris was hesitant to do this matrix inversion, because such an inversion was indeed messy\cite{sto02}.
However, in our operator formalism,
since $\C^2=-1$ when acting on vectors perpendicular to $\bh$, the above can
be inverted in a single line,
\ba
\bv_1&=&\frac{(1+\frac12\ta\C)}{(1+\ta^2/4)}[(1+\frac12 \ta\C)\bv_0+\dt\bac]\nn\\ 
&=&\frac{(1-\ta^2/4+\ta\C)\bv_0}{(1+\ta^2/4)}+\dt\frac{(1+\frac12\ta\C)\bac}{(1+\ta^2/4)}
\la{vone}
\ea
which is just (\ref{st3}) with Boris angle $\ta_B$. Thus Boris' original velocity 
update (\ref{vone}), is mathematically identical to the extended form of Buneman's update (\ref{st3}),
when the rotation angle is $\ta_B$.

In (\ref{borrot}), the Boris rotation is produced by the operator 
\be
{\cal R}_B=\frac{1+\frac12 \ta\C}{1-\frac12\ta\C}.
\ee
Since $\C$ plays the role of ``$\sqrt{-1}=i$'', ${\cal R}_B$ is the norm-preserving
Cayley\cite{he15,kna15} or Crank–Nicolson\cite{ric20} form, which is just
 the [1/1] Pad\'e approximate of $\exp(\ta\C)$.

The velocity update (\ref{vone}) is exactly the same as the splitting
\ba
\bv_{1/2}&=&\bv_0+\frac12\dt\bac,\nn\\
\bv_R&=&R(\br,\ta_B)[\bv_{1/2}],\la{mrot}\\
\bv_1&=&\bv_R+\frac12\dt\bac,
\la{fv}
\ea
where the final velocity is
\ba
\bv_1&=&R(\br,\ta_B)[\bv_0+\frac12\dt\bac]+\frac12\dt\bac\nn\\
&=&R(\br,\ta_B)[\bv_0]+\dt\frac12(R(\br,\ta_B)+1)[\bac]\la{oner}\\
&=&R(\br,\ta_B)[\bv_0]+\frac{(1+\frac12\ta\C)}{(1+\ta^2/4)}\dt\bac.
\la{spone}
\ea
The last equality follows only because the rotation in (\ref{oner}) is Boris rotation $R(\br,\ta_B)$.
In general, the splitting result (\ref{fv}) can only be a second-order approximation
to the exact result (\ref{st3}).
Since for a general $R(\br,\ta)$ at $\dt<<1$, 
\ba
\dt\frac12(R(\br,\ta)+1)[\bac]\rightarrow
\dt\left[\ta(\frac12-\frac{\ta^2}{12}+\cdots)\bh\times\bac+(1-\frac{\ta^2}{4}+\cdots)\bac\right],
\ea
(\ref{oner}) only agrees with (\ref{smalldt}) to second-order in $\dt$.

Thus the second-order Boris solver EB2B is then just the following EM2B algorithm
\ba
\br_{1/2}&=&\br_0+\frac12\dt \bv_0\nn\\
\bv_1&=&\bv_B(\br_{1/2},\bv_0,\dt)+\bv_F(\br_{1/2},\bv_0,\dt)\la{v2b}\\
\br_1&=&\br_{1/2}+\frac12 \dt \bv_1
\la{em2b}
\ea
with the rotation angle $\ta$ in (\ref{v2b}) replaced by $\ta_B$, so that now for
general vectors $\bv$ and $\bac$, 
\ba
\bv_B(\br,\bv,\dt)&=&\bv+\frac{\ta(\bh\times\bv)+(\ta^2/2)\bh\times(\bh\times\bv)}{1+\ta^2/4},\nn\\
\bv_F(\br,\bv,\dt)&=&\dt\left[\bac+\frac{(\ta/2)(\bh\times\bac)+(\ta^2/4)\bh\times(\bh\times\bac)}{1+\ta^2/4}\right].
\la{vfb}
\ea
One can easily check that for perpendicular $\bv$ and $\bac$, the above reduce to (\ref{vone}).
One is also free to replace $\bv_1$ above by the splitting form (\ref{fv}) with parallel components.
The splitting was a convenience in rotating the velocity vector only once, avoiding the
explicit form (\ref{vfb}), but not a necessity.
Boris solvers EB1B and EB1A correspond to updating the position first a full time step then the velocity and
vice versa. Again, EBLF is just EB1B with initial position $\br_0-\frac12\dt\bv_0$.

The Boris solver EB2B is unique in that: 1) it completely avoids the $1/\w$ singularity of
the $\E\times\B$ drift term, 2) its E-B splitting is exact, rather than just second-order 
approximation, and 3) its trajectory is on-orbit for a constant $\E$ and $\B$ field for $\dt$ of any size.  

\begin{figure}[hbt]	
	\includegraphics[width=0.49\linewidth]{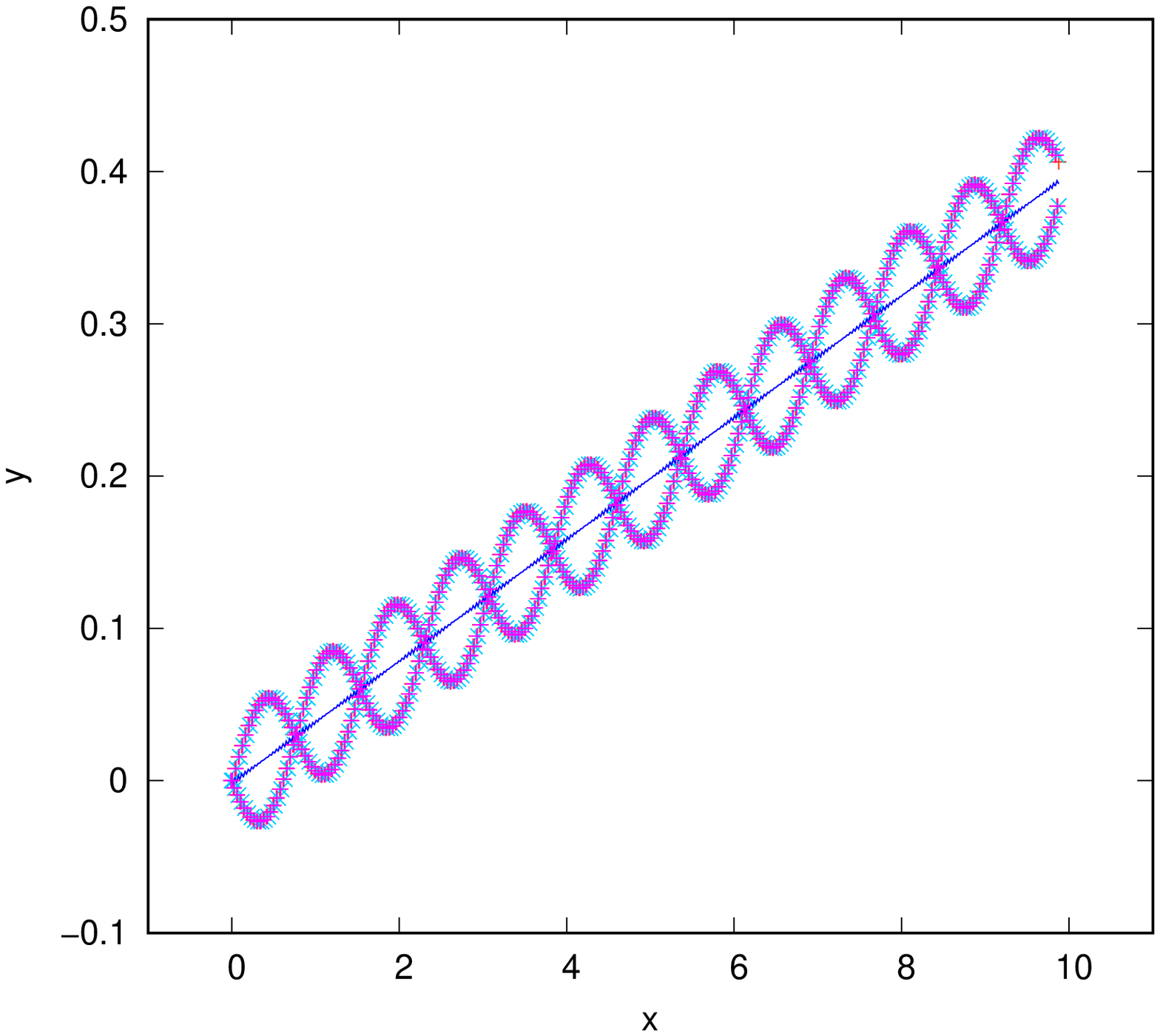}
	\includegraphics[width=0.49\linewidth]{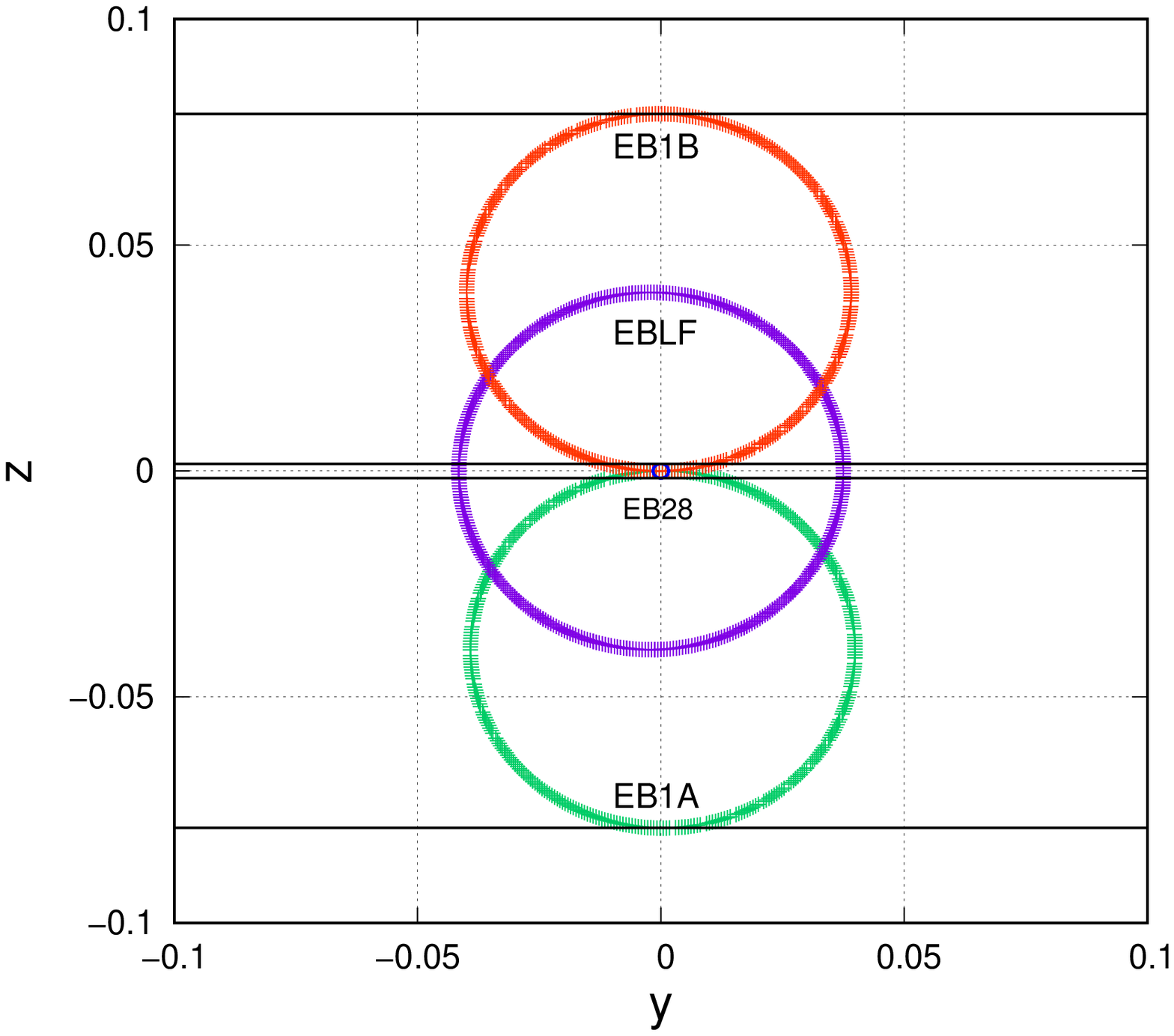}
	\caption{ (color online) {\bf Left} (a): $\E\times\B$ drift calculation using EB1A, 
		EB1B, EBLF and EB2B at $\dt=0.1975$.
		{\bf Right} (b): After removing the $\E\times\B$ drift in the y-direction, the oscillation in the
		$z$ coordinate reveals the gyrocenter location and gyroradius of all four solvers. EB2B is the tiny
		blue circle at the origin.
	}
	\la{pfone}
\end{figure}

In Fig.\ref{pfone}(a), we use EB1A, EB1B, EBLF and EB2B to compute an electron's trajectory in a 
combined $\E= \bk$ and $\B=250 \bi$ field with $\w=B=250$, $\bac=-\bk$, $\br_0={\bf 0}$ and $\bv_0=0.1\bi+0.4\bk$. 
Since the magnetic field predominates, the gyroradius is closely given by $r_g=0.4/\w=0.0016$, with
gyroperiod $T=2\pi/\w\approx 0.02513$. We needed $\dt=0.1975$, approximately eight times the
gyroperiod, to closely match the 13 first-order oscillations of Parker and Birdsall's\cite{par91} original Fig.1. No such large first-order oscillations are seen in EB2B's trajectory.
Fig.\ref{pfone}(a) is primarily used to verify the $\E\times\B$ drift velocity $v_d=1/\w=0.004\bj$, 
this drift can again be removed so that the erroneous $R_g$ of EB1A, EB1B and EBLF can be 
made manifest in Fig.\ref{pfone}(b).
Since $R_g\approx r_g\ta/2$ at large $\ta$ and the $y_c$ off-set is also 
$r_g\ta/2$, the maximum deviation for EB1A and EB1B from the true gyrocenter is 
$\pm r_g\ta=\pm 0.079$. This is the top most 
and bottom most horizontal lines in Fig.\ref{pfone} (b). By contrast, EB2B only oscillates near zero
within the true gyroradius $\pm 0.0016$.   

\begin{figure}[hbt]	
	\includegraphics[width=0.49\linewidth]{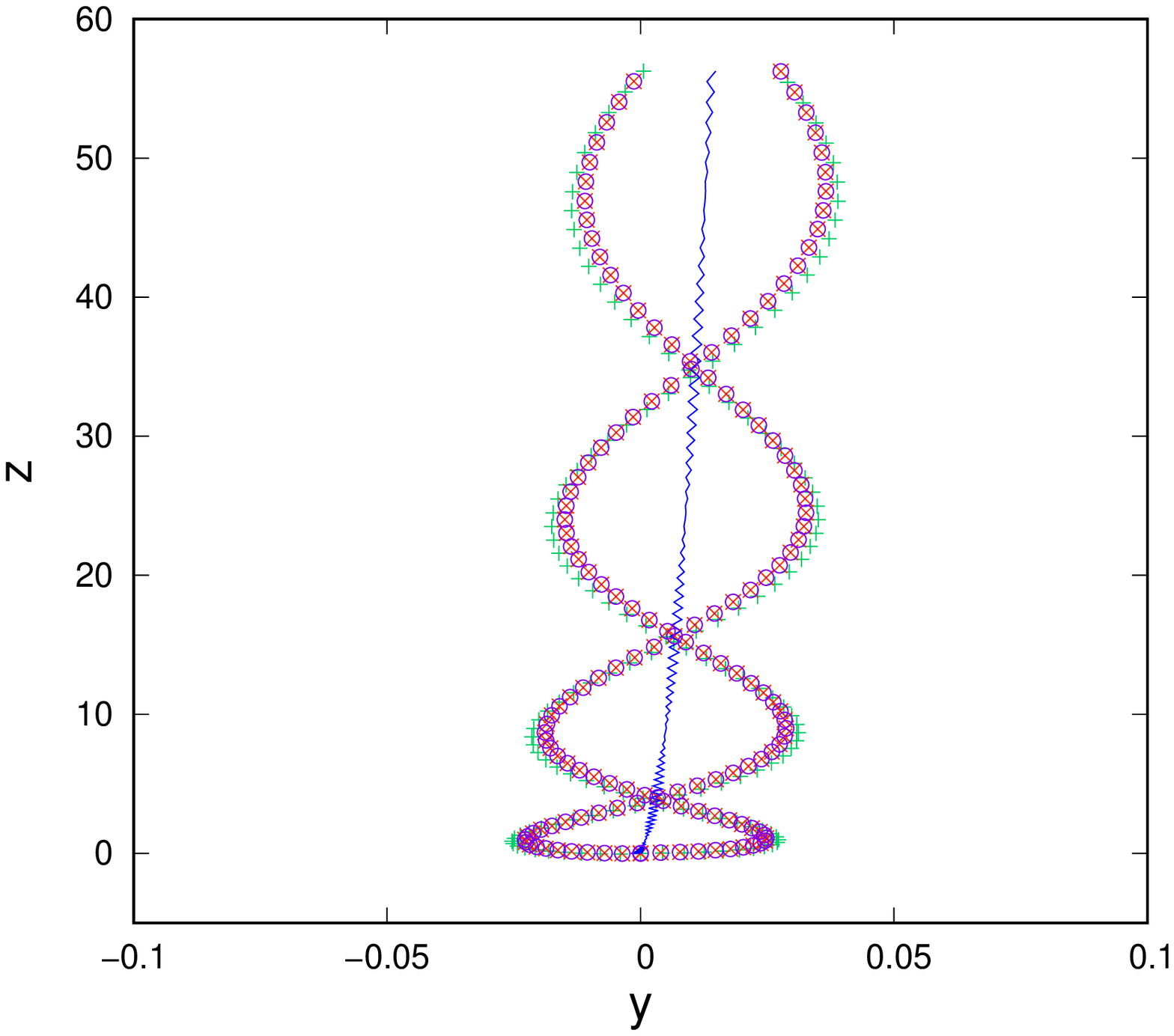}
	\includegraphics[width=0.49\linewidth]{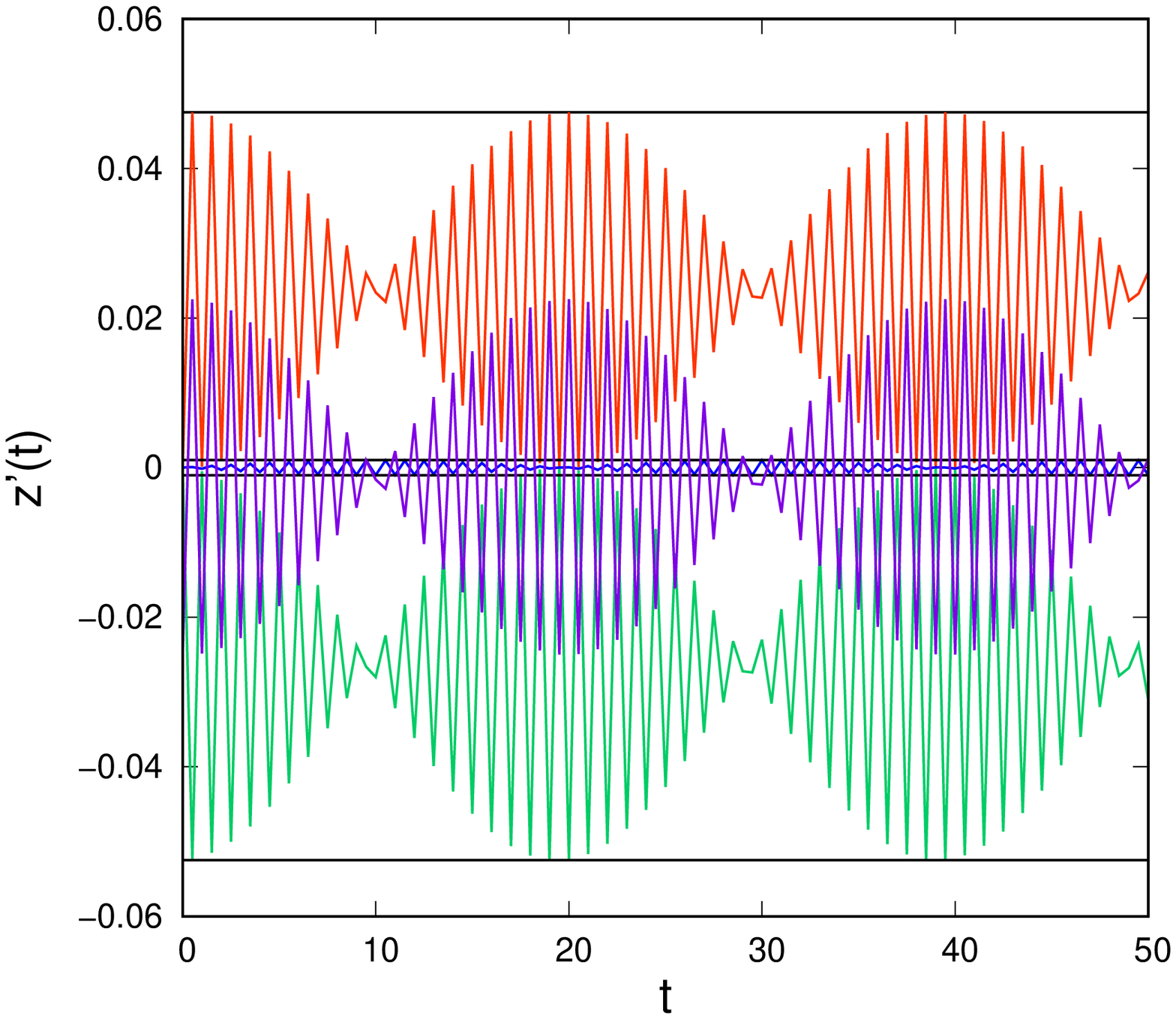}
	\caption{ (color online) {\bf Left} (a): Polarization drift calculation using EB1A (green +), 
		EB1B (red $\times$), EBLF (purple $\circ$) and EB2B (blue line) at $\dt=0.5$.
		{\bf Right} (b): The oscillation of $z'(t)=z(t)-0.01t^2$ for measuring gyro-diameters of all four Boris solvers. 
		Results for EB1B, EBLF, EB1A are given by the top red, middle purple and bottom green lines respectively.
		The tiny oscillating line near zero is that of EB2B.
	}
	\la{pffour}
\end{figure}
  
In Fig.\ref{pffour}(a) we use the same four solvers to reproduce Parker and Birdsall's\cite{par91}
Fig.4 on the polarization drift with $\E=(-2t)\bj$, $\B=100\bi$, $\br_0={\bf 0}$ and $\bv_0=0.1\bk$.
Again the large $R_g$ oscillations are absent from EB2B's trajectory.
Since the gyrocenter off-sets are in the direction of $\bv_0=0.1\bk$, they are not visible
along the $y$-coordinate. The $\E\times\B$ velocity drift here is $\bv_d=(2t)/B\, \bk$.
Removing the resulting coordinate drift gives $z'(t)=z(t)-0.01\,t^2$ which is plotted in Fig.\ref{pffour}(b). 
The top and the bottom lines are $0.04751$ and $-0.052467$ from zero, giving {\it different} gyro-diameters to EB1B and
and EB1A respectively. However, their average is correctly $0.04999=r_g\ta$. Since EBLF is nearly the same as EB1B,
its diameter is 0.04749. Its center is also slight off at $-0.0012$.
Again, EB2B's gyro-diameter 
showed no such $\dt$-dependence and is precisely bracketed by 
$\pm r_g=\pm 0.001$ around zero.

\begin{figure}[hbt]
	\includegraphics[width=0.80\linewidth]{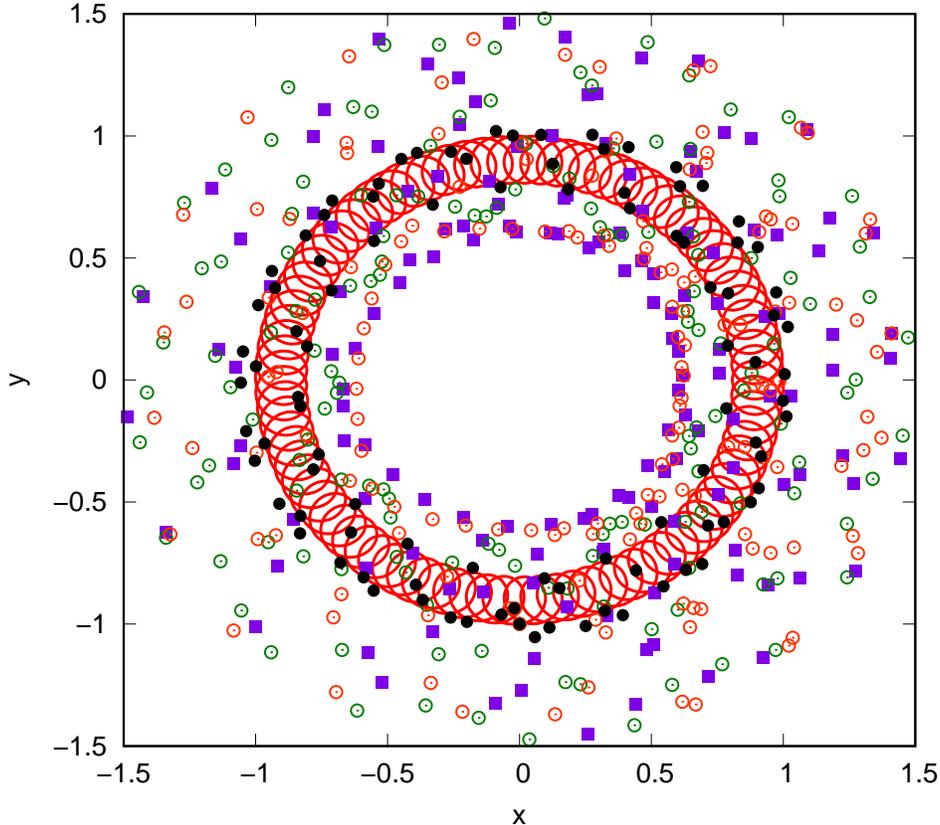}		
	\caption{ (color online)
		Open green and red circles are trajectory points of first-order Boris solvers EB1A and EB1B.
		Purple squares are those of leap frog Boris solver EBLF.
		Solid black circles are trajectory points of symmetric Boris solver EB2B. All are computed at
		$\dt=2.1\pi$. They are to be compared with 
		the solid red cycloid produced by EM2B at $\dt=\pi/10$. 
	}
	\la{ring}
\end{figure}

For a more modern comparison, with combined $\nabla\B$ and $\E\times\B$ drifts, 
the following 2D field configuration from Ref.\onlinecite{he15},
\be
\B=r\hat{\bf z}\qquad \bac=\frac{10^{-1}}{r^{3}}(x\hat{\bf x}+y\hat{\bf y})
\ee
with $r=\sqrt{x^2+y^2}$, is also tested. For $\br_0=(0,-1)$ and $\bv_0=(0.1,0.01)$,
the motion is a super-circle of gyro-circles with gyro-period $T=2\pi$. The trajectory
computed with EM2B at $\dt=T/20$ is shown as solid red line in Fig.\ref{ring}. 
Trajectory points of EB1A, EB1B, EBLF and EB2B at $\dt=2.1\pi>T$ are plotted without their 
distracting connecting lines. The trajectory of EB2B remains close to the exact solution 
while those of EB1A, EB1B and EBLF are widely scattered. 
All non-Boris integrators, such as
EM2B, are unbounded at such a large $\dt$.  

\section{Conclusions and future directions}
\la{con}

In this work, we have derived various Boris solvers on the basis of the Lie operator method,
the same formalism used to derive symplectic integrators. The advantage of this approach
is that it can uncover trajectory errors, which are the foundational basis for Boris solvers,
not obvious from finite-difference schemes. The glaring off-set error
of the gyro-center, as well as that of the gyro-radius, can be used to easily identify 
first-order or leap-frog Boris solvers in historical calculations and current discussions.

Our formalism provides a global view of the structure of algorithms, as illustrated in
Sect.\ref{diff}, which unambiguously differentiate the intrinsic symmetric second-order Boris solver
from the conventional leap frog Boris solver.
Such a global view naturally suggests a simple explanation of why their gyroradii 
are different, as shown in Fig.\ref{blfcomp}. This observation is not obvious from just
examining the analytical form of their respective algorithm.
       
By using the cross-product operator $\C$, we were able to show easily the equivalence of the
velocity update in Buneman's drift-subtracting scheme, Boris's original inversion algorithm,
and Boris' E-B splitting method. Most surprisingly, 
we found that Boris' E-B splitting was
unnecessary in that the there was no problem for the splitting to solve, 
when Buneman's drift-subtracting scheme is properly implemented as (\ref{nosp}).
The realization that one has effectively $\C=``i$'' immediately make many results obvious.
Representing the cross-product as a $3\times 3$ matrix\cite{sto02,he15,kna15} completely obscures
this crucial insight. 
  
By repeating some historical calculations, this work showed that the second-order Boris solver EB2B,
can be used for large $\dt$ calculations with far greater accuracy than previously thought. It is
the only algorithm currently known to be stable at $\dt$ greater than the local gyro-period
for nonuniform fields. The obvious future direction is to devise beyond second-order, more accurate
Boris-like integrators which are simultaneously stable at large time steps.  

\medskip

\noindent
{\bf Acknowledgment }

Our understanding of the leap frog algorithm, as summarized by Fig.\ref{lfdig},
was inspired by one of the Reviewer's comments on this work.

\noindent
{\bf Declaration of competing interest}

The authors declare that they have no known competing financial interests or personal relationships that could have appeared to influence the work reported in this paper.

\noindent
{\bf Data availability }

The data that supports the findings of this study are available from the corresponding author upon reasonable request.

\noindent
{\bf Funding}

This research did not receive any specific grant from funding agencies in the public, commercial, or not-for-profit sectors.

\appendix

\section {Two exact magnetic field solvers}
Appendix A of Ref.\onlinecite{chin08} has shown that two second-order algorithms 
for a constant magnetic field can be exactly on the gyro-circle if the updating
steps in (\ref{m2a}) are modified to
\ba
\bv_1&=&\bv_B(\br_0,\bv_0,\dt/2) \nn\\
\br_1 &=& \br_0+ \dt\Bigl[\bv_1+g(\theta)\bh\times(\bh\times\bv_1) \Bigr]\nn\\ 
\bv_2&=&\bv_B(\br_1,\bv_1,\dt/2) 
\la{m2ap}
\ea
and those in (\ref{m2b}) are modified to 
\ba
\br_1 &=& \br_0+ \frac12\dt\Bigl[\bv_0+h(\theta)\bh\times(\bh\times\bv_0) \Bigr]\nn\\ 
\bv_1&=&\bv_B(\br_1,\bv_0,\dt) \nn\\
\br_1 &=& \br_1+ \frac12\dt\Bigl[\bv_1+h(\theta)\bh\times(\bh\times\bv_1) \Bigr]
\la{m2bp}
\ea
with
\ba
g(\theta)&=&1-\frac{\sin(\theta/2)}{(\theta/2)}\la{gf}\\
h(\theta)&=&1-\frac{\tan(\theta/2)}{(\theta/2)}.
\la{hf}
\ea
The essence of the Boris solver is to decouple the rotation angles in (\ref{gf}) and (\ref{hf})
from $\ta$ to $\ta_C$ and $\ta_B$ such that $\sin(\ta_C/2)=\ta/2$ and $\tan(\ta_B/2)=\ta/2$,
forcing $g(\ta)=0=h(\ta)$ and (\ref{m2ap}) and (\ref{m2bp}) back to the form (\ref{m2a}) and (\ref{m2b}),
yielding solvers C2A and B2B.

In Appendix B of Ref.\onlinecite{chin08} the same two conditions $g(\ta)=0=h(\ta)$ also yield
exact trajectories in a constant electric and magnetic field.

\newpage

\end{document}